\documentclass{article}
\usepackage{graphicx} % Required for inserting images
\usepackage{hyperref} % para direcciones web
\counterwithout{table}{section}  % Desvincula el contador de tabla del contador de sección
\usepackage{adjustbox}
\usepackage[T1]{fontenc}
\usepackage{array}
\usepackage{float}
\usepackage{geometry}
\usepackage{subcaption}
\usepackage{caption}
\usepackage{tabularx}
\usepackage{rotating}
\usepackage{enumitem}
\usepackage{geometry}
\usepackage{array}
\usepackage{multirow}
\usepackage{booktabs}
\usepackage{amssymb}
\usepackage{longtable}
\usepackage{amsmath}
\usepackage[table,xcdraw]{xcolor}
\usepackage{rotating}
\usepackage{afterpage}
\usepackage{authblk}

\title{A systematic review of norm emergence in multi-agent systems}
\author[1,2]{Carmengelys Cordova}
\author[1,2]{Joaquin Taverner}
\author[1,2]{Elena Del Val}
\author[1,2]{Estefania Argente}

\affil[1]{Valencian Research Institute for Artificial Intelligence (VRAIN). Universitat Politècnica de València, Valencia, Spain}
\affil[2]{Valencian Graduate School and Research Network of Artificial Intelligence (valgrAI)}

\affil[ ]{\texttt{\{ccorgar3, joataap,  edelval, eargente\}@vrain.upv.es}}

\counterwithout{table}{section}
\begin{document}

\maketitle
\begin{abstract}
Multi-agent systems (MAS) have gained relevance in the field of artificial intelligence by offering tools for modelling complex environments where autonomous agents interact to achieve common or individual goals. In these systems, norms emerge as a fundamental component to regulate the behaviour of agents, promoting cooperation, coordination and conflict resolution. This article presents a systematic review, following the PRISMA method, on the emergence of norms in MAS, exploring the main mechanisms and factors that influence this process.
Sociological, structural, emotional and cognitive aspects that facilitate the creation, propagation and reinforcement of norms are addressed. The findings highlight the crucial role of social network topology, as well as the importance of emotions and shared values in the adoption and maintenance of norms. Furthermore, opportunities are identified for future research that more explicitly integrates emotional and ethical dynamics in the design of adaptive normative systems.
This work provides a comprehensive overview of the current state of research on norm emergence in MAS, serving as a basis for advancing the development of more efficient and flexible systems in artificial and real-world contexts.
  
\end{abstract}

\section{Introduction}
Multi-agent systems (MAS) represent an increasingly relevant approach within the field of artificial intelligence. These systems allow modelling complex environments in which autonomous agents interact to achieve common or individual goals. As these systems become more sophisticated, there is a need to establish mechanisms to regulate interactions between agents, ensuring cooperation, coordination and conflict resolution. In this context, norms become a crucial component for the regulation of behaviour in MAS, not only facilitating internal regulation, but also ensuring that agents' decisions are aligned with the overall expectations of the system.

The emergence of norms in MAS is a dynamic phenomenon, in which regulatory behaviours are not externally imposed, but emerge through interactions between agents. This spontaneous process of norm creation allows systems to be more flexible and adaptive to changing environments, replicating characteristics of human societies. Understanding how norms can emerge, diffuse and stabilise within these systems is essential for the design of agents that can act in a coherent manner, both in artificial environments and in real-world applications.

This article aims to conduct a systematic review, using the PRISMA method, that examines the current state of research on norm emergence in MAS. The review focuses on identifying the main mechanisms that facilitate this emergence process, with special attention to social, structural, emotional, cognitive and propagation factors that influence the norm emergence process.

The article is structured as follows: first, it reviews the emergence of norms from a sociological approach, exploring their emergence and evolution in human societies. Then, normative dynamics in societies of artificial agents are analysed, highlighting similarities and differences with human systems. Finally, a systematic review based on the PRISMA method is presented, synthesising the most relevant works on the mechanisms of norm emergence in MAS and their impact on the collective behaviour of agents.

\section{Theoretical Background in Human Societies}

\subsection{Normative approaches}
\label{sec:enfoques_normativos}

\begin{figure}[t]
  \centering
  \includegraphics[width=0.4\textwidth]{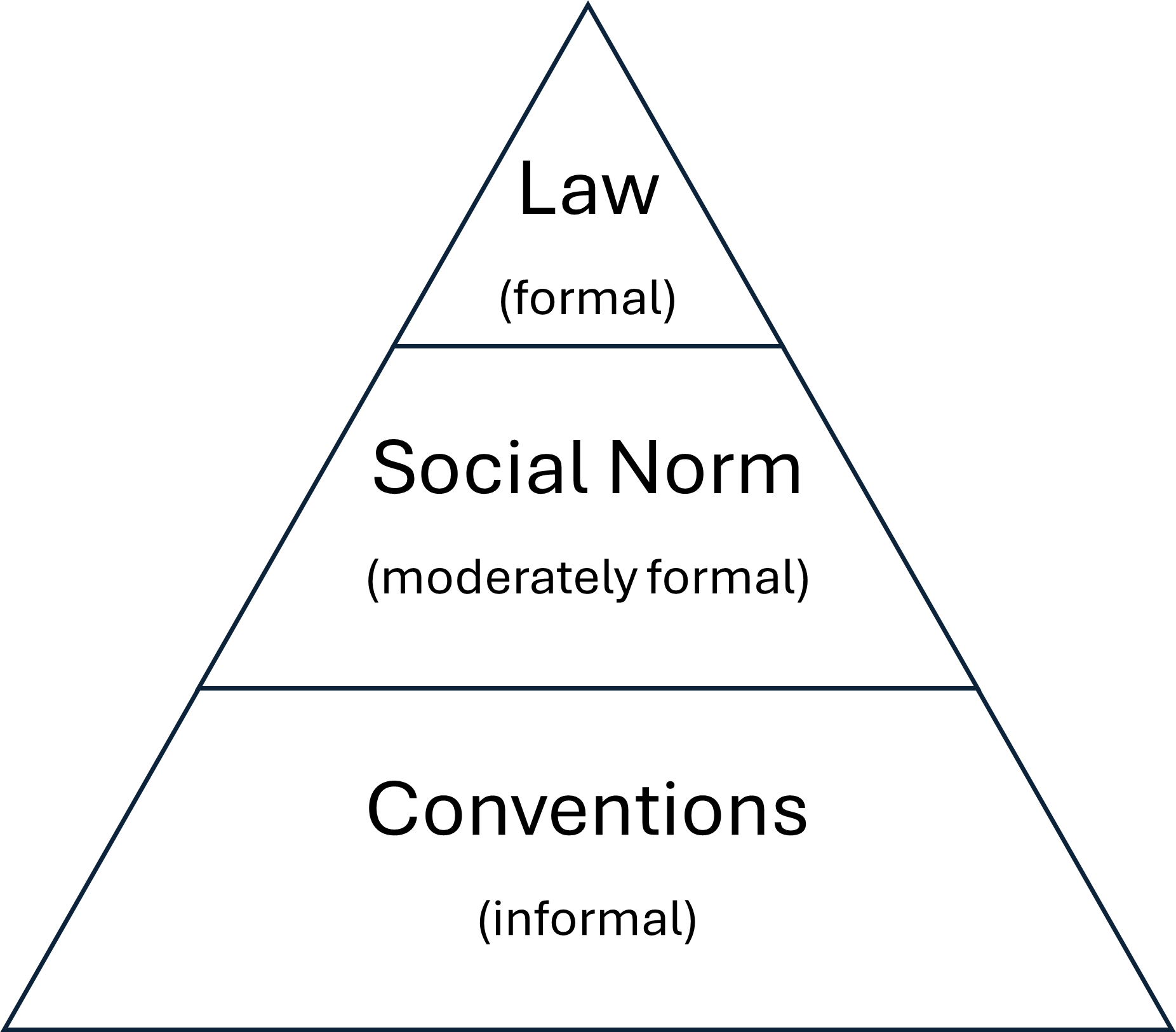}
  \caption{Categorisation of norms.}  
  \label{fig:piramidesocial}
\end{figure}

\begin{table}[t]
\centering
\caption{Comparison of the prescriptive and emergent perspectives in norm analysis.}
\label{table:norm_perspective}
\begin{adjustbox}{max width=\textwidth}
\renewcommand{\arraystretch}{1.5}
\begin{tabular}{p{5cm} p{5cm} p{5cm}}
\toprule
\textbf{Feature} & \textbf{Prescriptive Approach} & \textbf{Emergent Approach} \\
\midrule
\textbf{Origin} & Centralized imposition & Emergence through interactions \\
\textbf{Formalization} & High, explicitly codified norms & Low, implicitly learned norms \\
\textbf{Adoption} & Mandatory, authority-based & Voluntary, based on utility and benefits \\
\textbf{Flexibility and Adaptability} & Limited, stable and rigid norms & High, adaptable and evolving norms \\
\textbf{Sanctions} & Formal, clearly defined & Informal, based on social approval \\
\textbf{Example} & Traffic laws & Giving up a seat for elderly people on a bus \\
\bottomrule
\end{tabular}
\end{adjustbox}
\end{table}

The analysis of norms has been a widely debated topic in disciplines such as philosophy, sociology and law. Over the years, various approaches to understanding how norms that regulate behaviour within societies are established and perpetuated have been proposed  \cite{roig2021derechos}. These approaches can be broadly categorised into two distinct approaches: the prescriptive approach and the emergent approach (Table~\ref{table:norm_perspective}). On the one hand, the prescriptive approach, also known as the deontic approach, is deeply influenced by the philosophy of law and contractualism \cite{hart2012concept,austin1861province}. The theories encompassed under this approach follow a perspective according to which norms are imposed from a central authority or hierarchical structure. In this sense, norms seek to be a vehicle for encouraging compliance with pre-established norms, ensuring social order. This approach has been essential for the configuration of formal legal systems and normative structures, as they are characterised by a high formalisation of norms with clearly defined formal sanctions. However, this perspective faces limitations in terms of flexibility and adaptability, as norms are often stable and rigid, making it difficult to modify them in changing environments.

On the other hand, the emergent approach is based on sociological theories, such as conventionality and self-organisation \cite{lewis2008convention,epstein1996growing}. These theories argue that norms are not externally imposed, but arise spontaneously through interactions between individuals within a social group \cite{lewis2008convention,elias2000civilizing}. From this perspective, norms are seen as implicit agreements that evolve through adaptation, learning and the reiteration of accepted behaviours within the community. Unlike the prescriptive approach, emerging norms are not formalised or explicitly codified, but are implicitly learned and voluntarily adopted based on their perceived usefulness and benefits. This approach therefore offers greater flexibility and adaptability than the prescriptive approach, as norms can evolve and adjust as social circumstances change. In addition, sanctions are often informal, based on social approval or disapproval.

The combination of prescriptive and emergent approaches provides a more complete and nuanced understanding of the processes of norm formation and maintenance, which is essential for understanding the coexistence of rigid normative structures and more fluid conventions that regulate quotidian social interactions \cite{durkheim1922division,giddens1984constitution}.

\subsection{Norm typology}\label{sec:norm_typology}

\begin{table}[t]
\caption{Characteristics of different types of norms.}
\label{table:conventions_norms_laws}
\centering
\begin{adjustbox}{max width=\textwidth}
\renewcommand{\arraystretch}{1.5}
\begin{tabular}{p{3cm} p{4cm} p{4cm} p{4cm}}
\toprule
\small \textbf{Characteristic} & \small \textbf{Conventions} & \small \textbf{Social Norms} & \small \textbf{Laws} \\
\midrule
\small \textbf{Origin} & \small Spontaneous, emerging from individuals' interactions. & \small Social, based on values and principles. & \small Centralized imposition, legal authority. \\
\small \textbf{Formalization} & \small Informal, not codified. & \small Moderately formal, partially codified. & \small Formal, codified in legal documents. \\
\small \textbf{Sanctions} & \small Informal, such as social disapproval. & \small Social, loss of reputation and possible restrictions within the group. & \small Formal, such as fines and legal penalties. \\
\small \textbf{Adoption} & \small Generally adopted voluntarily, aligning with efficiency and adaptability within society. People follow them because they facilitate social interaction and are practical. & \small Adopted due to social pressure and alignment with collective goals and values. There is a stronger expectation of compliance and more severe social consequences for non-compliance. & \small Mandatory, imposed by legal mandate and authority. \\
\small \textbf{Example} & \small Greeting with a handshake. & \small Giving up a seat to elderly people on public transport. & \small The mandatory use of seatbelts in cars. \\
\small \textbf{Description} & \small It is a commonly accepted practice to greet someone with a handshake in many cultures. It is not written or legally required, but people are expected to follow it to be polite and respectful in a social context. & \small This is an implicit rule that society expects to be followed to show courtesy and respect for the elderly. Although there is no law requiring it, failing to comply can result in social disapproval. & \small This is a regulation established by the government and codified in the legal system. Failing to comply with this law can result in fines or penalties. \\
\bottomrule
\end{tabular}
\end{adjustbox}
\end{table}

Once the approaches that explain the origin and persistence of norms have been identified, it is necessary to analyse how they materialise in the various forms they take within societies. Norms are not homogeneous and vary according to context and function, taking different forms. Three general typologies are generally used in the literature to classify norms (Figure \ref{fig:piramidesocial}): conventions, social norms and laws. Each of these typologies represents different levels of formalisation and enforcement mechanisms. Table~\ref{table:conventions_norms_laws} shows a summary of the characteristics of the different types of norms.

Social conventions are a set of norms, rules, or expectations that govern the behaviour of people within a society or group \cite{bicchieri2009right,rimal2005behaviors}. It can be understood as a tacitly admitted practice, arising from precedents or customs \cite{conte1999conventions}, establishing patterns of behaviour that emerge naturally without imposing deontological obligations \cite{savarimuthu2011norm,haynes2017engineering}. These conventions are understood and accepted by community members, which facilitates the organisation of social interaction, coexistence and cooperation between people. The absence of formal sanctions for the violation of conventions is what differentiates them from other forms of norms regulated by stricter collective expectations \cite{gibbs1965norms}.

Social conventions can vary widely among different cultures and societies, and can include such things as norms of dress, forms of greeting, and appropriate ways of speaking or acting in public \cite{markus2014culture}. These norms are not always written or formally established, but are taught and transmitted through socialisation. A clear example of this is the handshake as a form of greeting, a common practice in many Western cultures.  Although there is no legal obligation to do so, most people do it to respect others and to preserve social order and harmony. Conventions can be transformed into social norms, especially when society values conformity and is willing to punish those who do not conform to the expectation of social behaviour. Thus, a pattern of behaviour that initially begins as a convention may come to be regarded as a social norm \cite{southwood2011norms}.  For example, in Japanese culture, bowing, which originally emerged as an act of respect and humility in ancient Japanese courts, has become a deeply rooted social norm. Most people feel an obligation to follow this gesture as a sign of respect, especially in formal situations. Ignoring or improperly performing the bow can be interpreted as a lack of respect and generate deep disapproval.

A social norm, unlike a convention, imposes an explicit obligation on how individuals should act within a society. Non-compliance can result in social sanctions, although not formal ones, which are effective in regulating behaviour \cite{savarimuthu2009norm, haynes2017engineering}. These sanctions include disapproval, rejection, or even deliberate exclusion of an individual or group from a society. These mechanisms ensure conformity in order to preserve social cohesion \cite{fehr2004social}. Social norms, therefore, operate as informal regulators, which depend not only on widespread acceptance, but also on the community's willingness to punish offenders. Punishment is not only imposed by  those directly affected by the violation, but may also be carried out by members outside the group, a phenomenon known as ‘altruistic punishment’ \cite{fehr2004third}. This type of punishment, which seeks to protect the collective welfare, reinforces the social norm and promotes conformity within society \cite{bicchieri2006grammar}.  This is key in groups where cooperation and interdependence are essential, as it allows norms to be maintained and evolve to reflect the needs of the group, strengthening cohesion and trust within the community. Furthermore, it is important to note that social norms are not static, but adapt over time to new social, economic or technological circumstances. Social norms can therefore vary significantly across cultures and historical contexts, underlining their dynamic nature \cite{mahoney2009explaining}.

Finally, a law \cite{haynes2017engineering} is a rule established by a higher authority to regulate, in accordance with justice, some aspect of social relations. In human societies, it is common for many social norms to be formalised in a set of legal rules. A norm becomes law when there is the possibility of a sanction or punishment to ensure compliance. These actions are enforced by persons specifically charged with enforcing the law, due to their authority or position \cite{gibbs1965norms}. For example, social norms prohibiting violence are often codified in laws with specific sanctions.

\subsection{Life cycle of a social norm}
\label{sec:live_cycle}

\begin{table}[t]
\centering
\caption{Summary of approaches on the life cycle of social norms. Note that N/A represents that the phases not used in that model.}
\label{table:live_cycle}
\begin{adjustbox}{max width=\textwidth}
\renewcommand{\arraystretch}{1.5}
\begin{tabular}{p{1.5cm} p{2.0cm} p{3.5cm} p{3.0cm} p{2.4cm} p{3.0cm}}
\toprule
\small \textbf{Proposal} & \small \textbf{Phase 1: Creation} & \small \textbf{Phase 2: Diffusion and Adoption} & \small \textbf{Phase 3: Internalization} & \small \textbf{Phase 4: Forgetting} & \small \textbf{Phase 5: Transformation} \\
\midrule
\small \cite{rogers1962teoria} & \small Knowledge & \small Persuasion &  \small Decision \& Implementation & Confirmation & \small N/A \\
\small \cite{elster1989cement} & \small Creation & \small Sanctions \& Compliance & \small N/A &  \small N/A & \small Change \& Adjustment \\
\small \cite{axelrod1986evolutionary} & \small Emergence & \multicolumn{2}{c}{\small Stabilization and Internalization}  & \multicolumn{2}{c}{\small Change/Disappearance}  \\
\small \cite{coleman1990foundations} & \small Emergence & \small Adoption & \small Internalization \& Consolidation & \small Disappearance & \small Evolution \& Maintenance \\
\small \cite{finnemore1998international} & \small Emergence & \small Cascading Effect & \small Internalization & \small N/A & \small N/A \\
\small \cite{bicchieri2006grammar} & \small Emergence & \multicolumn{2}{c}{\small Stabilization} & \small N/A & \small Destabilization \& Change \\
\small \cite{fligstein2015theory} & \small Emergence & \small Consolidation & \small N/A & \small N/A & \small Transformation/Crises \\
\small \cite{wiener2018contestation} & \small Constitution & \small Diffusion \& Adoption & \small Contestation & \small N/A & \small Transformation \& Maintenance or Reconstitution \\
\bottomrule
\end{tabular}
\end{adjustbox}
\end{table}

Social norms regulate human behaviour independently of a formal legal framework and are stabilised through social pressure and community sanctions, emerging spontaneously in contexts where regulation arises from social interaction. In this sense, the life cycle of a social norm reflects its dynamic nature and its capacity to adapt to social transformations.  

The life cycle of a norm encompasses both its creation and acceptance and its ability to adapt or disappear in response to social change. Different models for defining the life cycle of norms have been proposed \cite{bicchieri2005grammar}. Table~\ref{table:live_cycle} shows a summary of the different proposals for the life cycle of a social norm. In general, the life cycle of a social norm can be described by five phases: creation, diffusion, internalisation, forgetting and transformation. The first phase in the life cycle of a norm is the norm creation phase. This process can be initiated deliberately by normative entrepreneurs \cite{finnemore1998international} or social actors seeking to solve common problems or collective needs \cite{coleman1990foundations}. Emergence can also be the result of more spontaneous social interactions within a community \cite{axelrod1986evolutionary,bicchieri2006grammar}.

In the second stage, the norm is disseminated and begins to be adopted by more individuals or actors within a social context. This stage is characterised by a process in which actors become aware of the norm and begin to evaluate it. This results in a cascade effect in which the norm is quickly adopted by a critical number of actors \cite{finnemore1998international}, consolidating itself through mechanisms such as imitation and/or punishment of violators \cite{axelrod1986evolutionary}. The norm eventually becomes institutionalised within the community as it becomes stabilised and alienated with the dominant behaviour \cite{fligstein2015theory,bicchieri2006grammar}. In addition, this phase is also defined by the global adoption of the norm by a wide range of actors, which facilitates its diffusion and consolidation \cite{wiener2018contestation}.

During the third phase of the life cycle, individuals begin to internalise the norm, which means that it is no longer seen as an external imposition and becomes a part of their regular behaviour, becoming part of their values and beliefs \cite{coleman1990foundations}. As a result, compliance with the norm becomes less dependent on external sanctions, as individuals act in accordance with it according to their personal conviction and moral values \cite{finnemore1998international,rogers1962teoria,elster1989cement}. 

The fourth phase represents the extinction of the norm. Norms can disappear if they are no longer useful or if the social conditions that sustained them change \cite{axelrod1986evolutionary}. It has also been argued that norms can enter into crisis and disappear when the balances of power in a social field change \cite{fligstein2015theory}. 

Finally, in the fifth phase, norms experience destabilisation and change when social expectations are modified \cite{bicchieri2006grammar,fligstein2015theory}. In contrast to the fourth phase, norms do not disappear, but go through a process of maintenance or reconstitution \cite{wiener2018contestation,coleman1990foundations}. This phase implies that norms are challenged, but survive and adapt to new consensuses or global realities, managing to be reconstituted or transformed in order to continue to be in practice.

\subsection{Factors influencing the emergence of social norms}
 \label{sec:emergency_factors}

\begin{figure}[t]
  \centering
  \includegraphics[trim = 8mm 35mm 8mm 35mm, clip, width=\textwidth]{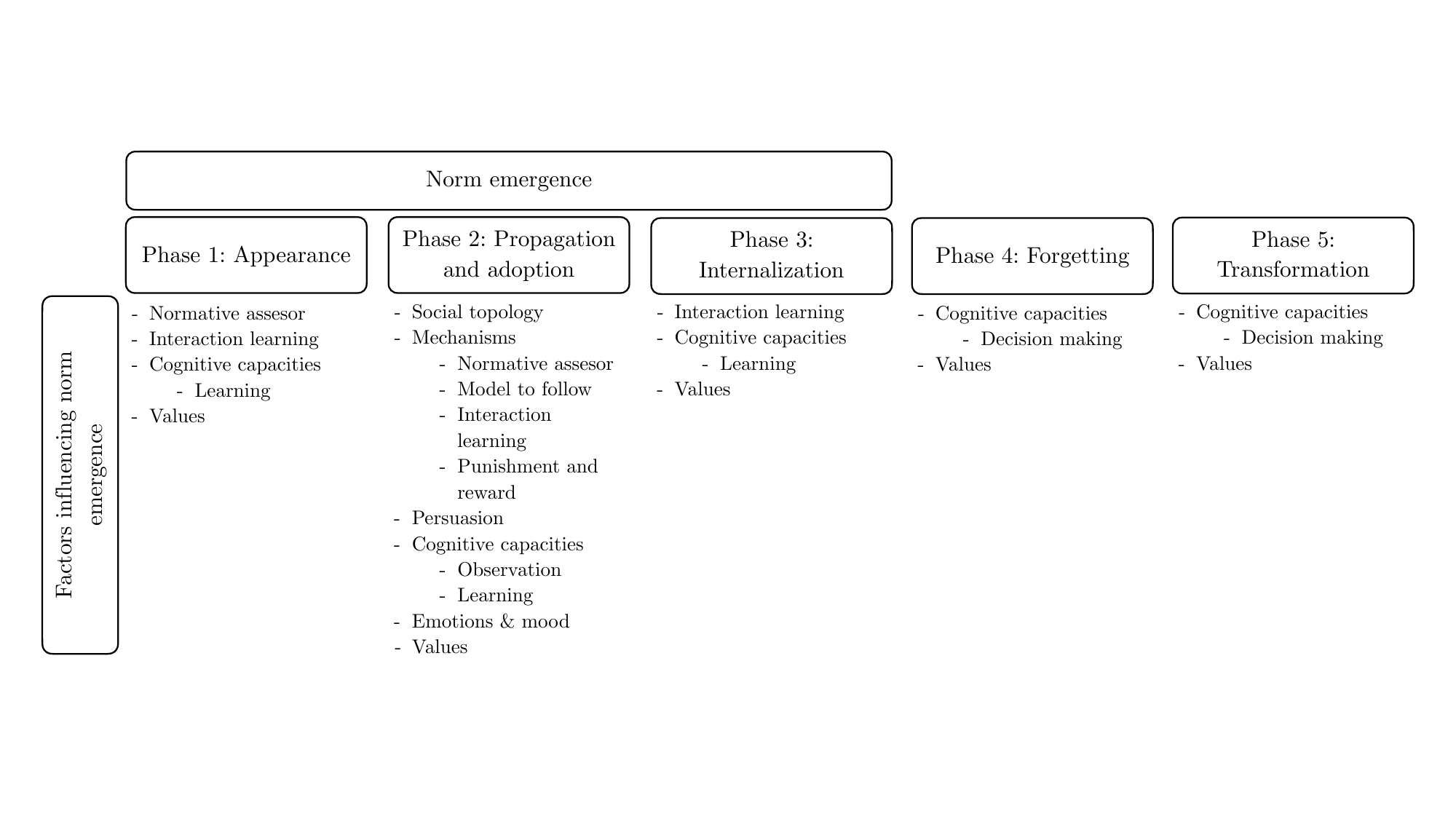}
  \caption{Factors influencing the social emergence of a norm. }  
  \label{fig:emergency_factors}
\end{figure}

Among the phases of the life cycle of a norm discussed in the previous section, the emergence phase is the most relevant for the purpose of this paper. As mentioned above, the emergence of a norm is a phenomenon that arises from interactions between members of a society. This process encompasses the first three phases of the life cycle described in the previous section (see Figure~\ref{fig:emergency_factors}): creation, diffusion and internalisation. Understanding the mechanisms that trigger the emergence of norms is essential to understanding how these norms are developed and established in a society \cite{bicchieri2016norms}. However, the emergence of a norm often occurs in a non-standardised and decentralised way, making it a process that is difficult to predict \cite{holland2000emergence,prokopenko2009information}.

The emergence of a norm is a complex process in which reaching a significant level of consensus is critical. For a norm to be considered emergent within a community, it is necessary for a significant percentage of individuals to adopt and follow it consistently. This process involves not only individual acceptance, but also a shared expectation that other members of the community will comply with the norm \cite{trujillo2021norm,andrighetto2015perceived,zhang2023we}. Therefore, the stability of emerging norms is important to avoid oscillation between normative behaviours, which could lead to uncertainty and make it difficult to consolidate stable collective behaviours. Structural and cultural factors, such as social topology, propagation and adoption mechanisms, individuals' cognitive capacities, emotions and values, influence the stability of norms and the process of emergence as detailed below. 

\subsubsection{Social typology}
\label{sec:social_typology}
Social topology refers to the structure of relationships and connections between individuals within a social network. This arrangement not only determines how actors interact, but also influences the propagation and adoption of behaviours and social norms \cite{granovetter1973strength,coleman1990foundations}. In a social network, the nodes represent individuals or actors, while the links or connections between them represent interactions or relationships. The configuration of these connections determines, to a large extent, how norms emerge and stabilise in a community. Various factors in the social topology play a crucial role in the emergence of norms: 

\begin{enumerate}[label=\roman*)]
    \item \textbf{Network diameter}: defined as the greatest distance between two individuals, it is a key factor in determining the speed at which norms propagate. Networks with smaller diameters tend to have greater cohesion between individuals, which accelerates the diffusion of common norms \cite{watts1998collective}. This is because shorter distances allow interactions to occur more quickly and allow norms to be transmitted more efficiently between individuals.
    \item \textbf{Neighbourhood size}: in a social network, this refers to the number of direct connections an individual has with other individuals. This factor is important for the propagation of norms, as a larger neighbourhood increases awareness of diverse normative influences and facilitates their diffusion. Individuals with more direct connections often play a key role in the adoption and stability of norms within the network, due to the social pressure exerted by their frequent interactions \cite{valente1996social}.
    \item \textbf{ Clustering coefficient}: measures the likelihood that an individual's neighbours are connected to each other. Networks with a high level of clustering tend to form cohesive subgroups where local norms can be more easily established. These subgroups can function as micro-communities within the wider network, in which norms are developed and maintained through social surveillance and sanctioning \cite{jackson2008social}.
    \item \textbf{ Intermediation centrality (or betweenness)}: describes the ability of certain individuals to act as bridges between different parts of the network. Individuals with high betweenness are key to the diffusion of norms, as they connect subgroups that would otherwise be disconnected from the network. In this way, norms that emerge in one subgroup can be transmitted to other parts of the network, encouraging the adoption of large-scale normative behaviour \cite{freeman1978centrality}.
    \item \textbf{ Network density}: measures the overall proportion of existing connections compared to all possible connections. A dense network, in which most individuals are connected to each other, promotes greater social cohesion and facilitates the emergence of cooperative norms. This is because, in highly dense networks, individuals are more interconnected and therefore norms are more likely to be reinforced through surveillance and social sanctions \cite{reagans2001networks}. 
    \item \textbf{Weak ties}: defined as less frequent or close connections between individuals, they play a crucial role in the propagation of norms across different groups within a network. Although these ties may seem less significant, they are instrumental in connecting distant parts of the network and allowing norms to be diffused beyond the immediate subgroups \cite{centola2007complex}. Weak ties allow emerging norms to be propagated between separate communities, resulting in a wider and more diverse adoption of normative behaviours.
\end{enumerate}

\subsubsection{Social propagation mechanisms}
\label{sec:propagation} 
The mechanisms of norm propagation and adoption facilitate the dissemination and adoption of norms by a wide group within a society \cite{centola2007complex}. The propagation of norms is also connected to the idea of internalisation. Through the repetition of normative interactions, individuals come to internalise these norms, incorporating them into their everyday practices without the need for external reinforcement \cite{bourdieu1977outline}. This suggests that, once norms have been effectively propagated, their adoption becomes a social habit that automatically guides individuals' behaviour. There are several mechanisms of norm propagation:

\begin{enumerate}[label=\roman*)]
\item \textbf{Normative advisor}: in human societies is defined as an individual, who plays a key role in the propagation, interpretation and stabilisation of social norms \cite{frantz2014modelling}. His or her main function lies in facilitating the understanding, internalisation and adoption of norms, both formal and informal, by acting as a mediator between collective expectations and individual behaviour. This role involves transmitting norms through mechanisms such as persuasion, example and education, as well as monitoring compliance and evaluating their impact on social cohesion and stability. Normative advisors also participate in the process of redefining norms in contexts of social change, contributing to their evolution and adaptation \cite{finnemore1998international}
\item \textbf{Role model}: is a figure, such as a celebrity, community leader, influencer, or any person with a significant and respected presence, who serves as an example to others. By observing how these role models adopt certain norms, other members of the community are motivated to imitate their behaviour. The influence of these role models is based on the admiration and respect they generate, which is particularly effective in societies where social conformity and imitation of prominent figures are important values.
\item \textbf{Learning by interaction}: is based on the idea that norms are transmitted and consolidated through daily social interaction. Individuals learn and adopt new norms by observing and participating in activities with other members of their community. This learning process can be informal, (i.e., in everyday conversations, observations and shared experiences) or structured (i.e., educational or work environments where normative behaviour is actively encouraged). The repetition and consistency in these interactions help internalise the norm, making it part of individuals' habitual behaviour.
\item \textbf{Punishments and rewards}: are relevant for facilitating compliance and ensuring social cohesion. Punishments, which can range from warnings to stricter measures, are essential to provide a dissuasive incentive for non-compliance and reinforce the importance of following the norms \cite{hofer2019efficacy}. Effective punishment underlines the relevance of the norm and ensures that individuals understand the negative consequences of non-compliance. On the other hand, rewards motivate individuals to adopt the norms voluntarily, creating an environment where compliance is perceived as advantageous \cite{fehr2000fairness,henrich2001people}. Therefore, the balanced application of punishments and rewards is a key element in maintaining normative order, as it promotes not only compliance, but also the perceived legitimacy of norms in society \cite{mulder2010rules,tyler2006people}.
\end{enumerate}

In addition to the factors previously discussed, there are other factors that complement the propagation and adoption of norms in a society. For example, persuasion directly influences individual decisions through convincing arguments or incentives that facilitate the adoption of social behaviours \cite{axelrod1986evolutionary}. Another influencing factor is social pressure, as it plays a crucial role in imposing collective expectations, leading individuals to follow the norms in order to avoid being sanctioned or \cite{cialdini2004social}. Finally, external factors such as inequality or insecurity can also play a role as they can intensify the need for shared norms, reinforcing compliance with norms \cite{wilkinson2010spirit}. As a result of these factors, conformity emerges, i.e. people adopt socially accepted behaviours in order to better integrate into the community \cite{asch1956studies}.

\subsubsection{Cognitive abilities}
Cognitive capacities, both individual and collective, also influence the emergence of norms in human societies. These capacities encompass from moral and strategic reasoning to shared cognition in social groups, enabling people to interpret, create and modify norms that regulate their behaviour and contribute to the maintenance of social order \cite{cokely2009cognitive}. At the individual level, these skills enable individuals to process information, reason and make decisions, facilitating the internalisation and adaptation of social norms. This process involves the interpretation of social expectations and the ability to adjust behaviour to established norms. The emergence of norms is also linked to the ability to anticipate the expectations of others, which requires skills such as strategic reasoning, understanding and predicting the beliefs and intentions of others \cite{bicchieri2005grammar}.

From a collective perspective, norms emerge through the joint processing of information, which highlights the importance of cognitive coordination between individuals \cite{hutchins1995cognition}. Cognitive coordination refers to the process by which several people synchronise their mental capacities and knowledge to achieve a common goal, enabling collaborative information processing and shared decision-making. Thus, norms are not only based on individual capacities, but also on the interaction and cooperation of collective cognitive capacities \cite{kant1964groundwork}. 

At a general level, three main factors can be identified concerning the cognitive capacities that make the propagation of norms possible: observation, learning and decision-making. First, observation is a fundamental cognitive ability that enables individuals not only to perceive, but also to interpret and understand the behaviour of others in their environment \cite{stone2000multiagent}. This ability is key for individuals to detect patterns of behaviour that they can imitate or avoid, depending on the observed outcomes of those actions. Observation influences both imitation and collective learning, becoming the basis for individuals to adopt or reject norms within their society \cite{bandura1977social}. 

Second, learning enables individuals to adapt to their environment by the acquisition of knowledge, skills and behaviours based on experience \cite{cosmides1994beyond}. In the context of human societies, learning occurs not only at the individual level, but also socially, through observation and interaction with other members of the community. The learning process facilitates the internalisation of collectively accepted behaviours, contributing to the cohesion and stability of social norms \cite{bandura1977social,tomasello1993cultural}. Moreover, the ability to learn from experience and adapt to new situations allows individuals not only to follow established norms, but also to actively participate in their creation and modification in response to contextual changes. Thus, learning becomes a driver of normative change, promoting the evolution of social norms that govern collective behaviour \cite{boyd1988culture,piaget1976piaget}.

Finally, through the ability to make decisions, individuals choose specific actions based on available information, previous learning and socially internalised norms. This process involves not only the consideration of personal preferences, but also the evaluation of social expectations and consequences, which reinforces the coherence of behaviour within a group or community \cite{kahneman2013prospect,ajzen1991theory,kahneman2011thinking}. 

\subsubsection{Emotions}

Emotions directly influence the behaviour of individuals within a group. Emotions, as automatic and rapid responses to specific stimuli or situations, not only affect immediate interactions between people, but also determine how individuals perceive and adapt to social expectations \cite{lerner2015emotion}. Although short-lived, the intensity of emotions can be key to creating or reinforcing norms. Visible emotional responses, such as facial expressions or physiological changes, allow other members of the community to adjust their behaviour in line with emotional cues, thus facilitating the spread of accepted norms \cite{cohn2009positive}. For example, in \cite{tangney2007moral}, shame motivates agents to conform to group norms to avoid social disapproval, which contributes to normative cohesion and stability. 

Another relevant aspect is people's ability to anticipate future emotions based on their actions or decisions. The expectation of negative emotions, such as guilt or regret, may prevent individuals from breaking norms, while the anticipation of positive emotions, such as pride or gratitude, motivates conformity to norms \cite{mellers2001anticipated, patrick2009affective}. This process of emotional projection is essential for the stability of norms, as people, by anticipating the emotional consequences of their actions, adjust their behaviour in a way that reinforces normativity in the social group \cite{perugini2001role, steenhaut2006mediating}. On the other hand, empirical evidence also supports the idea that emotions can influence the speed and direction of normative change. In \cite{van2015persuasive} it has been shown that when negative emotions are strong and widely shared, they can accelerate the process of normative change, while positive emotions consolidate emerging norms by reinforcing desired behaviours. Thus, emotions not only act as signalling and regulatory mechanisms that allow individuals to identify, internalise and react to norms, but are also essential for group cohesion and social adaptation.

\subsubsection{Values}

Finally, values, understood as ethical and moral principles that guide human behaviour, lay the foundations that structure any society. Values act as underlying criteria that determine which behaviours are acceptable or unacceptable. They therefore play a crucial role in social cohesion and in the emergence of norms that guide the behaviour of individuals within a community. In this context, norms emerge as tools that institutionalise these values, facilitating cooperation and mutual respect between individuals. Values, shared by the members of a society, serve as a frame of reference that guides their collective actions and decisions \cite{durkheim1984division}.  On the other hand, norms do not arise in isolation, but are deeply linked to the prevailing values of a society. When these values are consensual and recognised by the majority, the derived norms tend to be more widely accepted \cite{habermas1985theory}. For example, values such as fairness, respect and justice are fundamental for norms to reflect principles of equality of opportunity and fair treatment for all individuals \cite{rawls1971theory}. 

In societies where values are fragmented or in conflict, norms can lose their legitimacy \cite{macintyre2007after}. If there is no clear consensus on the fundamental principles guiding the community, norms risk being perceived as ineffective or unfair, which can lead to social tensions and lack of compliance. The legitimacy of norms therefore depends on their ability to reflect and promote the shared values of society.

\section{Normative Multi-agent Systems}
\label{sec:nmas}

Multi-agent systems (MAS) is an interdisciplinary branch of artificial intelligence that focuses on the design and analysis of systems composed of multiple agents capable of interacting with each other \cite{bond2014readings,wooldridge2009introduction}. An agent is an autonomous entity capable of perceiving its environment, making decisions and performing actions to fulfil its goals \cite{taverner2017gestion}. Interaction between agents, whether cooperative or competitive, is fundamental in MASs, as it allows complex problems to be solved through task distribution and coordination. To achieve this, agents must adapt and learn from both the environment and their interactions \cite{russell2016artificial}.

The development of agent behaviour often involves the use of metaphors that try to simulate human aspects such as reasoning or social behaviour. One of the relevant aspects in the simulation of agent societies is the use of norms or rules to regulate interactions between agents, ensuring cooperation, coordination and conflict resolution. Over the years, numerous studies have evaluated the advantages and disadvantages of using both explicit and implicit norms to regulate the behaviour of NMAS. This work falls under the umbrella of normative MAS (NMAS) \cite{viana2021towards,dell2020runtime,hollander2011current}. An NMAS can be defined as a set of agents whose interactions are regulated by norms that determine permitted, prohibited or obligatory behaviours \cite{jones2001handbook, argente2020normative}. Generally, this type of system allows the norms and their effects on the agent's reasoning or decision-making process to be represented to a greater or lesser extent \cite{mahala2023normative,garcia2009constraint}. Many of the existing NMAS also include mechanisms and tools to detect both compliance and violation of such norms \cite{boella2008substantive, mahmoud2014review}.

\subsection{Previous surveys}
The emergence of norms in NMAS has been a topic covered in previous reviews. For example, in \cite{savarimuthu2011norm}, the authors conducted a review of the state of the art in NMAS focusing on the treatment of the life cycle of norms. The results of the analysis emphasised the strengths and weaknesses of each of the mechanisms studied and identified the main characteristics when modelling the life cycle of norms in NMAS. Another interesting proposal can be found in \cite{haynes2017engineering}. In that study, a review of the main proposals dealing with the use of emerging norms and social conventions in NMAS was carried out. To do so, they studied the mechanisms proposed in the literature covering three key stages of norm emergence: detection, evaluation, and diffusion. Finally, the most recent work analysing proposals in NMAS is from 2019 and was presented in \cite{morris2019norm}. That paper analysed and classified the main proposals in the field of standards emergence. 

The review presented in this article completes previous reviews with an updated analysis of the existing literature following the PRISMA method, providing a systematic and rigorous evaluation of the current state of the art. This review has been approached from the perspective of analysing the process of norm emergence in human societies as a theoretical framework for a subsequent analysis of work on the process of norm emergence in the area of NMAS. For each of the phases of the emergence process, the factors that influence the emergence of norms have been analysed, as well as the mechanisms used for their simulation in NMAS. Among these factors, the analysis of the mechanisms used in the literature to simulate the effect of emotions and social values on the emergence of norms represents a novelty in the field. 

\subsection{Systematic review}

\begin{figure}[t]
  \centering
  \includegraphics[width=0.8\textwidth]{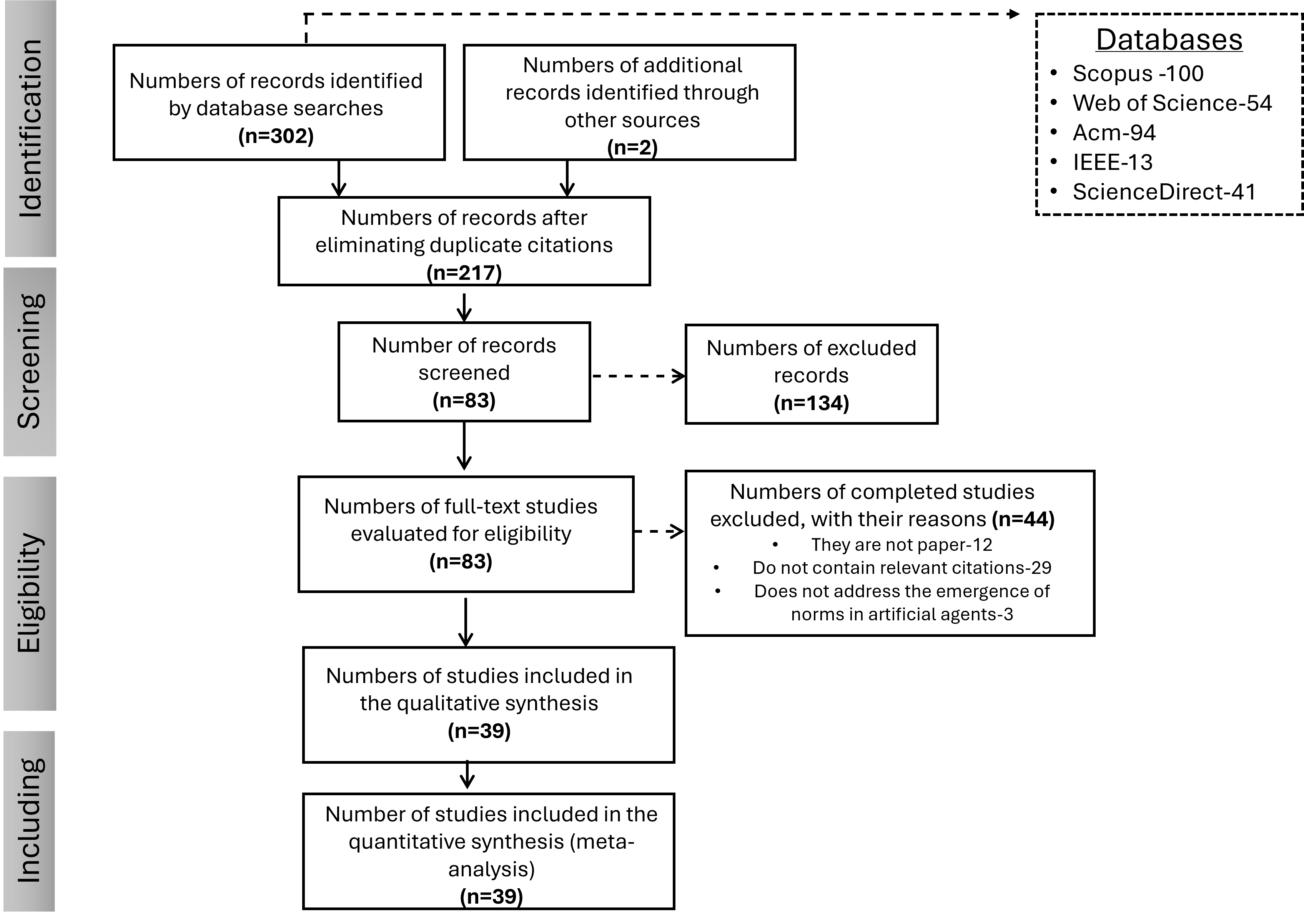}
  \caption{Prisma flowchart.}  
  \label{fig:prisma}
\end{figure}

The present review was conducted following the Preferred Reporting Items for Systematic Reviews and Meta-Analyses (PRISMA) methodology \cite{molins2019bases,moher2009preferred,page2021prisma}. This methodology is based on a series of structured steps that include the precise definition of inclusion and exclusion criteria, an exhaustive literature search in multiple databases, the critical evaluation of the quality of the selected studies and, finally, the rigorous synthesis of the results. Figure~\ref{fig:prisma} shows the process followed during the systematic review. 

The search was conducted in March 2024 and the results were limited to publications from 2005 (inclusive) to the present. The query combined the following terms: (`Norm emergence' OR `emergence') OR (`Normative systems’') AND (`Artificial Intelligence' OR `Computing Intelligence' OR `Agent'). The search results yielded a total of 302 articles: 100 in Scopus, 54 in Web of Science, 94 in ACM, 13 in IEEE, and 41 in ScienceDirect. The need to broaden the review by hand searching to ensure that no relevant papers were omitted was considered. To this end, Google Scholar was searched using the aforementioned search terms. As a result of this search, two additional articles were included, \cite{frantz2014modelling,ren2024emergence}, which were considered fundamental to the analysis due to their relevance and depth on the topic.

\begin{table}[t]
\centering
\caption{Inclusion and exclusion criteria for selecting studies to review.}
\label{table:criteriosPRISMA}
\begin{adjustbox}{max width=\textwidth}
\renewcommand{\arraystretch}{1.5} % Incrementa el espacio entre filas
\begin{tabular}{p{7cm} p{7cm}}
\toprule
\textbf{Inclusion Criteria} & \textbf{Exclusion Criteria} \\
\midrule
Research whose title and abstract are clearly related to the topic of norm emergence in multi-agent systems. & Studies focusing on norms pre-established by humans or institutions, without analyzing norm emergence from agent interactions. \\
Studies investigating how norms emerge in multi-agent systems. & Studies not focused on computational and artificial intelligence methods as tools for norm emergence. \\
Studies focusing on artificial intelligence as a tool for norm emergence. & \\
Studies using computational methods to model or simulate norm emergence in agents or multi-agent systems. & \\
Studies employing empirical methods to analyze norm emergence in real-world systems. & \\
\bottomrule
\end{tabular}
\end{adjustbox}
\end{table}

The initial systematic review covered a total of 304 articles, encompassing a period from 2005 to 2024, providing a comprehensive and up-to-date overview of the state of research in the field of multi-agent system standards. In order to make the selection of the articles that finally form part of the review, duplicates were removed from the 87, leaving 217 articles for review. The remaining 217 articles were then thoroughly screened, applying the established inclusion and exclusion criteria (see Table~\ref{table:criteriosPRISMA}). Titles and abstracts were then reviewed and only those papers that addressed the emergence of norms in MAS were selected, leaving a total of 83 articles. Specifically, 12 of these articles contained the relevant terms in the title only, 47 in the abstract only, and 24 in both the title and abstract, thus meeting the inclusion criteria. The next stage consisted of an in-depth analysis of the 83 articles. This resulted in the exclusion of 44 articles: 12 of them were excluded because they were proceedings books and not academic articles. In addition, 29 articles were removed because, although they were related thematically, they did not meet the citation relevance criterion. This criterion is assessed on the basis of a ratio (R) that establishes the relationship between the year of publication and the number of citations obtained: 

\begin{equation}
R = \frac{C}{2024-A}
\end{equation}

\noindent where $(C)$ is the number of citations obtained and $(A)$ is the publication date of the article. Articles with a ratio of less than 1 ($R<1$) were removed, as it was considered that they did not have a sufficient number of citations per year. Recent articles (years 2023 and 2024) were automatically included in the review regardless of the number of citations. Finally, 3 articles were excluded because they did not specifically address the emergence of norms in MAS, focusing on aspects that were not relevant to the objective of our research. After this process, 39 articles that met all the inclusion criteria were selected to be part of the review.

\subsection{Approaches to norms in MAS}
Reviewing the proposals included in this study, it is possible to identify three different types of approaches to deal with the emergence of norms: prescriptive approach, emergent approach and hybrid approach. On the one hand, the prescriptive approach in NMAS \cite{savarimuthu2011emergence}, also known as ``norms as prescriptions''. \cite{conte1999conventions}, is characterised by the centralisation of the emergence of norms. This approach usually involves a higher authority that defines and regulates the norms or rules that actors must follow \cite{morris2019norm}. This superior entity defines a set of guidelines and mechanisms for adherence and compliance to ensure the coherence and efficiency of the system \cite{savarimuthu2011norm}.

In general, this perspective usually refers to the use of deontic logic, i.e., norms are developed under concepts such as permissions, prohibitions, or obligations. Several works have proposed architectures for the study of norms, which encompass the creation of frameworks for describing and modelling norms using deontic logic in MAS \cite{cliffe2007embedding,aldewereld2010operetta,boissier2016jacamo,boella2006architecture, garcia2006norm,jones1993characterisation,wieringa1993applications}. These models also include the development of norm-reasoning agents in their decision-making process. For example in \cite{criado2010normative}, a NMAS for irrigation water management is proposed. The agent model is based on a BDI (Belief-Desire-Intention) architecture and the norms are pre-established in the NSMA itself. By using the norms the agents decide which irrigation policy maximises the quality of the crops. Another example of the use of explicit norms can be found in \cite{dastani2019classification}. The authors propose a BDI agent model capable of reasoning according to a set of pre-established obligations. 

On the other hand, in the emergent approach, norms are seen as preferred behaviours that develop spontaneously through interactions between agents. In that approach, emerging norms are observed as specific behaviours, strategies or policies selected from a set of possible actions in similar situations. Work in NMAS focuses on how norms emerge and diffuse within a society of agents, exploring different mechanisms such as leadership, reputation, learning and imitation \cite{ren2024emergence,abeywickrama2023emergence,mashayekhi2022prosocial,liu2021local}. Generally, in the emergent approach norms are implicit, i.e. they are not formally represented. The propagation of norms takes place through social interaction until a certain threshold of adoption is reached \cite{morris2021norm}.

For example, in \cite{savarimuthu2007mechanisms}, it is analysed how norms emerge in an NMAS through mechanisms such as imitation and reinforcement learning. A model of an agent capable of adapting its behaviour by observing the social consequences of its actions and those of other agents was proposed. Another interesting example can be found in \cite{yu2013emergence}. This work studies how social norms emerge within complex networks through collective learning in NMAS. To this end, the influence of interactions between agents and the configuration of the social topology on the formation of norms in a decentralised and spontaneous manner is analysed. Thus, the emergent approach facilitates the dynamic adaptation of the system, allowing norms to emerge on the basis of social behaviour. However, the implicit representation of norms can lead to confusion among agents as to which norms prevail in the system, especially if different subsets of agents adopt different norms.  

Finally, the hybrid approach combines elements of the prescriptive and emergent approaches, aiming to take advantage of the strengths of both. This approach combines the centralised model of norms with decentralised emergence, i.e. in some cases the social norm will emerge as a shared belief while in others it will be transmitted by an authority \cite{tzeng2022fleur,tzeng2022noe,macanovic2024signals}. For instance, work in \cite{ajmeri2018robust} addresses how norms emerge through a balance between the emergence of centralised norms and the ability of agents to reason about their adoption depending on the context. By using a hybrid approach, norms are allowed to be flexible enough to adapt to specific situations, while retaining the coherence needed to maintain stability in the system. The work of \cite{mashayekhi2016silk} proposes the ``Silk'' framework, which generates norms based on a central observer who identifies conflicts and suggests norms to avoid them. However, the decision to adopt or reject these norms depends on autonomous interactions between agents. Another normative framework for MAS is presented in \cite{morris2021norm}. In this framework, a set of agents, called synthesisers, generate and discuss norms based on requests from another set of agents, called participants. The acceptance of the norms is done through a decision mechanism. 

\begin{table}[t]
\centering
\caption{Classification of works into emergent and hybrid approaches.}
\label{tabla:enfoques}
\renewcommand{\arraystretch}{1.5} % Incrementa el espacio entre filas
\begin{adjustbox}{max width=\textwidth}
\begin{tabular}{p{7cm} p{7cm}}
\toprule
\textbf{Works with Emergent Approach} & \textbf{Works with Hybrid Approach} \\
\midrule
\cite{ren2024emergence,abeywickrama2023emergence,mashayekhi2022prosocial,liu2021local,mahmoud2017establishing,hao2017efficient,yu2017collective,mahmoud2016cooperation, yu2016adaptive,mahmoud2015establishing,yang2016accelerating,mukherjee2007emergence,savarimuthu2007mechanisms,savarimuthu2009norm,mukherjee2008norm,savarimuthu2007role,savarimuthu2009social,andrighetto2010complex,sen2009effects,savarimuthu2011aspects, brooks2011modeling,hollander2011using, mahmoud2012efficient, riveret2012probabilistic,yu2013emergence,yu2014collective,shibusawa2014norm} & \cite{tzeng2022fleur,tzeng2022noe,macanovic2024signals,levy2023convention,morris2021norm,ajmeri2018robust,mashayekhi2016silk,yu2015hierarchical,frantz2018modeling,campenni2009normal,beheshti2014normative,beheshti2015cognitive} \\
\bottomrule
\end{tabular}
\end{adjustbox}
\end{table}

\subsection{Representation of a norm in NMAS}

In the literature, it is possible to find two types of representation of norms: explicit and implicit. On the one hand, explicit norms are defined in a direct, formal and accessible way to all the agents of the system \cite{boella2003bdi,dos2013developing,morales2018off,conte1999conventions}. For example, in \cite{beheshti2014normative} a BDI normative agent model is presented that incorporates norms as part of the agent's internal knowledge in the form of beliefs. Other work, such as the one presented in \cite{lee2014n}, explores the use of an external, centralised representation within the NMAS. This representation acts as a normative knowledge base that is accessible to all agents in the NMAS \cite{cliffe2007embedding,hubner2009normative}. Generally, models that use explicit representations typically rely on the use of deontic logic by grouping norms into categories such as obligation, permission or prohibition \cite{mashayekhi2022prosocial,criado2011open}. A common way to regulate and monitor compliance with norms is through the use of control systems that simulate the action of a sanctioning authority \cite{djahel2014communications,al2015intelligent}. For example, the proposal made in \cite{badii2020smart} presents an agent model for traffic simulation in cities. The system has a set of driving norms, such as respecting traffic signs or speed limits. A central system checks compliance with the norms in real time and applies the appropriate sanctions. Another example can be seen in \cite{yu2015hierarchical}, where norms are implemented through guiding policies set by supervisors, who are in charge of synthesising information and directing agents' behaviour towards compliance in the hierarchical framework.

In contrast, implicit norms in multi-agent systems refer to norms or behavioural patterns that are not formally codified or explicitly defined within the system. During interactions, agents adjust their behaviour based on observation of each other's actions and implicit social expectations \cite{savarimuthu2011aspects}. For instance, \cite{mukherjee2007norm}, implicit social norms emerge through local and repeated interactions between agents distributed on a grid. These interactions, influenced by physical and private proximity between agents, allow norms to emerge spontaneously as preferred coordination solutions without centralised prescription. A further approach can be found in \cite{hollander2011current,hollander2011using}, where the authors propose a normative agent model in which norms are expressed implicitly. In that approach, agents develop their most effective responses or actions through their interactions. 

In addition to interaction, for a system to be able to simulate the emergence of norms, processes must be developed to maintain socially accepted behaviours over time, such as adoption systems, social pressure or indirect incentives. To address this internalisation process, in \cite{beheshti2014normative} a normative agent model was proposed that through social learning and BDI reasoning, the agent incorporated normative behaviours into its own cognitive structure. In that approach norms were approached as socially accepted behaviours. Through the agent's interaction, the agent gradually and progressively inferred which behaviours were accepted.

\subsection{Types of norms in NMAS}

\begin{table}[t]
\centering
\caption{Classification of works using an explicit representation of norms according to norm types: r-norms, i-norms, s-norms and m-norms.}
\label{tabla:normas_explicit}
\renewcommand{\arraystretch}{1.2} 
\begin{adjustbox}{max width=\textwidth}
\begin{tabular}{c c c c c}
\toprule
\textbf{Work} & \textbf{R-norms} & \textbf{I-norms} & \textbf{S-norms} & \textbf{M-norms} \\
\midrule
\cite{savarimuthu2007mechanisms} &  & $\checkmark$ & $\checkmark$ & $\checkmark$  \\ 
\cite{savarimuthu2007role} &  & $\checkmark$ & $\checkmark$ &  $\checkmark$ \\ 
\cite{andrighetto2010complex} &  & $\checkmark$ & $\checkmark$ &  $\checkmark$  \\ 
\cite{sen2009effects} & $\checkmark$ & $\checkmark$ & $\checkmark$ &  \\ 
\cite{campenni2009normal} &  & $\checkmark$ & $\checkmark$ &  $\checkmark$  \\ 
\cite{mahmoud2015establishing} & $\checkmark$ & $\checkmark$ & $\checkmark$ &  $\checkmark$  \\ 
\cite{mashayekhi2016silk} & $\checkmark$ &  &  & $\checkmark$  \\ 
\cite{yang2016accelerating} &  & $\checkmark$ & $\checkmark$ &  $\checkmark$  \\ 
\cite{yu2016adaptive} &  &  & $\checkmark$ &  \\ 
\cite{mahmoud2017establishing} &  & $\checkmark$ & $\checkmark$ &  $\checkmark$  \\ 
\cite{hao2017efficient} &  & $\checkmark$ & $\checkmark$ &  $\checkmark$  \\ 
\cite{yu2017collective} &  & $\checkmark$ & $\checkmark$ &  $\checkmark$  \\ 
\cite{morris2021norm} & $\checkmark$ &  &  & $\checkmark$  \\ 
\cite{frantz2018modeling} & $\checkmark$ & $\checkmark$ & $\checkmark$ & $\checkmark$  \\
\bottomrule
\end{tabular}
\end{adjustbox}
\end{table}

\begin{table}[t]
\centering
\caption{Classification of works using an implicit representation of norms according to norm types: r-norms, i-norms, s-norms and m-norms. }
\label{tabla:normas_implicit}
\renewcommand{\arraystretch}{1.2} 
\begin{adjustbox}{max width=\textwidth}
\begin{tabular}{c c c c c}
\toprule
\textbf{Work} & \textbf{R-norms} & \textbf{I-norms} & \textbf{S-norms} & \textbf{M-norms} \\
\midrule
\cite{mukherjee2007emergence} &  & $\checkmark$ & $\checkmark$ &  \\ 
\cite{savarimuthu2009norm} &  & $\checkmark$ & $\checkmark$ & $\checkmark$  \\ 
\cite{mukherjee2008norm} &  & $\checkmark$ & $\checkmark$ &  \\ 
\cite{savarimuthu2009social} &  & $\checkmark$ & $\checkmark$ &  \\ 
\cite{brooks2011modeling} &  & $\checkmark$ & $\checkmark$ &  \\ 
\cite{hollander2011using} &  & $\checkmark$ & $\checkmark$ &  \\ 
\cite{mahmoud2012efficient} &  & $\checkmark$ & $\checkmark$ &  \\ 
\cite{riveret2012probabilistic} &  &  & $\checkmark$ &  \\ 
\cite{yu2013emergence} &  & $\checkmark$ & $\checkmark$ &  \\ 
\cite{beheshti2014normative} &  & $\checkmark$ & $\checkmark$ &  \\ 
\cite{yu2014collective} &  & $\checkmark$ & $\checkmark$ &  \\ 
\cite{shibusawa2014norm} &  & $\checkmark$ & $\checkmark$ &  \\ 
\cite{beheshti2015cognitive} &  &  & $\checkmark$ &  \\ 
\cite{yu2015hierarchical} &  & $\checkmark$ & $\checkmark$ &  \\ 
\cite{mashayekhi2016silk} & $\checkmark$ &  &  &  \\ 
\cite{mahmoud2016cooperation} &  &  & $\checkmark$ &  \\ 
\cite{yu2016adaptive} &  &  & $\checkmark$ &  \\ 
\cite{ajmeri2018robust} &  &  & $\checkmark$ &  \\ 
\cite{tzeng2021noe} &  &  &  &  \\ 
\cite{tzeng2022fleur} &  & $\checkmark$ & $\checkmark$ &  \\ 
\cite{abeywickrama2023emergence} &  & $\checkmark$ & $\checkmark$ &  \\ 
\cite{levy2023convention} &  & $\checkmark$ & $\checkmark$ &  \\ 
\cite{ren2024emergence} &  & $\checkmark$ & $\checkmark$ & $\checkmark$  \\ 
\cite{macanovic2024signals} &  & $\checkmark$ & $\checkmark$ &  \\ 
\bottomrule
\end{tabular}
\end{adjustbox}
\end{table}

As seen in Section~\ref{sec:enfoques_normativos}, in human societies a distinction is made between conventions, social norms and laws. In NMAS, these norms have been classified in a similar way (see Figure 11), distinguishing between: institutional norms (r-norms), social norms (s-norms), interaction norms (i-norms) and moral or personal norms (m-norms). Table~\ref{tabla:normas_explicit} shows a classification of works using an explicit representation and different types of norms. In contrast, Table~\ref{tabla:normas_implicit} shows a classification of works using an implicit representation and different types of norms.
 
\begin{figure}[t]
  \centering
  \includegraphics[width=0.6\textwidth]{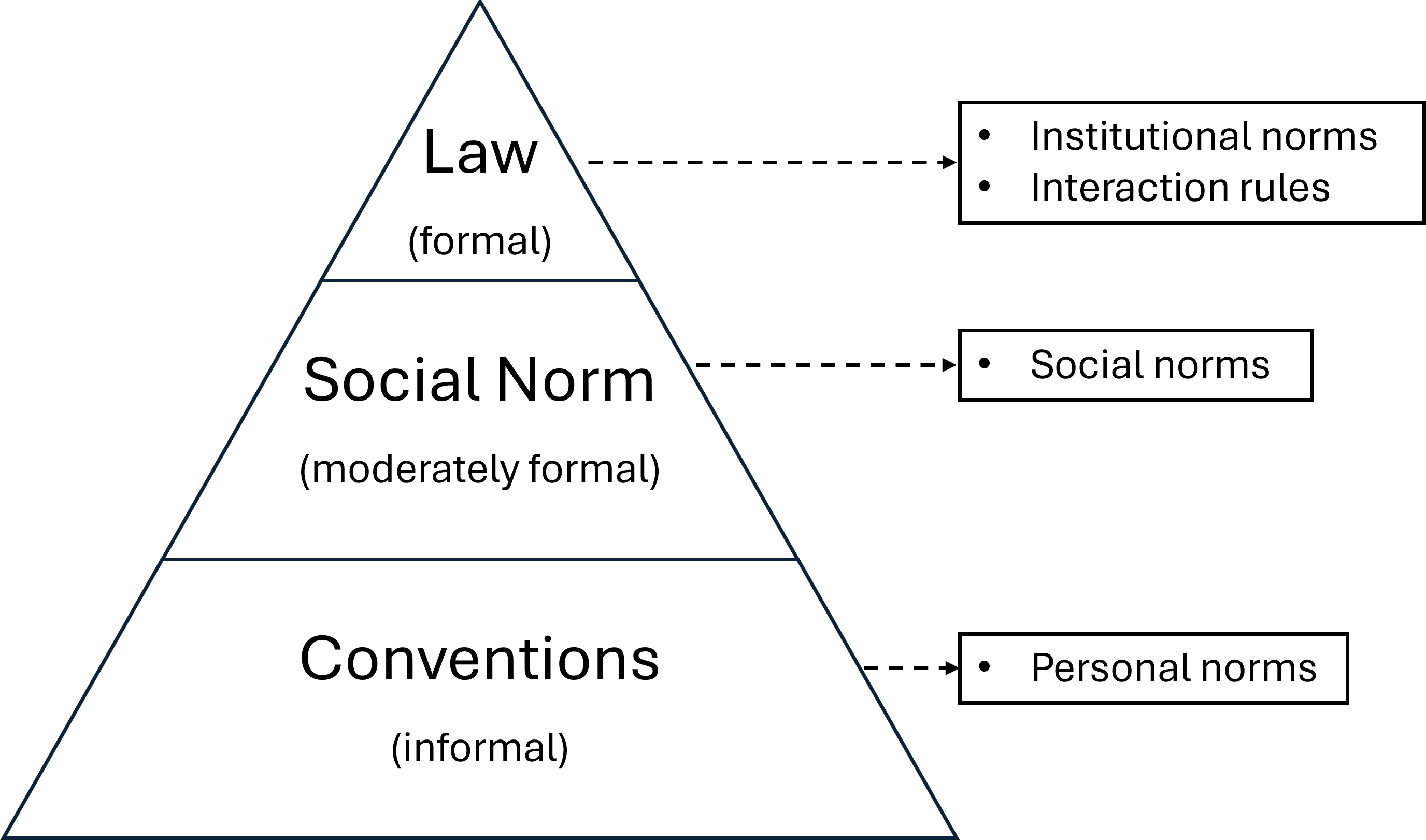}
  \caption{Categorización de las normas en MAS}  
  \label{fig:Figura11}
\end{figure}

Institutional norms are related to the concept of law, i.e. they represent guidelines formally established, communicated and enforced by regulatory authorities. One approach to simulating institutional norms in SMA is by centralising them in a single agent acting as a legislator. In the work presented in \cite{mashayekhi2016silk}, the use of a norm-generating agent acting as a central legislator is proposed. This agent dictates two types of laws or norms: strict norms, which prevent undesirable events; and flexible norms, which promote desirable behaviours and prevent conflicts. In addition, the generating agent is responsible for monitoring compliance with the norms in the interactions between the agents in the system. Another approach to the emergence of institutional norms is proposed in \cite{morris2021norm}. In contrast to the previous approach, the authors present a model based on the consensus of a group of agents to establish institutional norms. For this purpose, two types of agents are used, the participants and the synthesisers. On the one hand, participants identify norm needs through interaction. These needs are transmitted to the synthesising agents who act as legislators. These agents analyse the proposal and determine the need for a norm through a polling process. When a norm is created, the synthesising agents disseminate them so that participating agents can adopt them.    

Interaction norms simulate social norms to a certain degree \cite{balke2013norms}. For example, in \cite{mukherjee2007emergence} a bottom-up social learning model is proposed that allows emergent norms depending on interactions between agents. A scenario based on interaction between near neighbours is used. During the interaction, interaction norms emerge in each neighbourhood and spread to the rest of the neighbourhoods. For emergence to occur in this type of system it is necessary to establish mechanisms that allow agents to reach a convergence in social behaviour. In this sense, in \cite{brooks2011modeling} a mathematical model is proposed to achieve convergence in the emergence of interaction norms. To validate the model, a game in which agents must achieve convergence in their behaviour was used. The results of the experiments showed that the model was able to faithfully simulate the process of convergence in norm emergence. Similarly, in \cite{yang2016accelerating} an approach based on hierarchical heuristic learning is presented to simulate norm emergence in NMAS. This method organises learning in hierarchical layers, allowing agents to initially focus on high-level strategies before refining their behaviour with more specific norms. Experimental results showed that the model was able to improve agents' ability to adopt norms compared to traditional methods. In addition, the hierarchy-based algorithm reduces the cost of communication between agents while facilitating scalability. 

Like interaction norms, social norms also emerge organically from interactions between agents, seeking to reach consensus on socially accepted behaviours \cite{tuomela1995importance,savarimuthu2011norm}. These norms do not require explicit formalisation and represent expectations of behaviour collectively supported by a group of agents. The complexity of these norms lies in their emergent nature, as they are generally not explicitly described in society. Individuals must analyse their own and others' behaviour, as well as the consequences of these behaviours, in order to establish socially accepted \cite{argente2020normative} patterns of behaviour. As an example of social norms, \cite{sen2007emergence} uses the resolution of social dilemmas at traffic intersections. Agents must learn norms such as ‘yield to the vehicle on the right’ or ‘yield to the vehicle on the left’ in situations where both drivers arrive simultaneously. Other works such as the one developed in \cite{yu2013emergence} use the concept of a virtual society in which agents must decide whether to drive on the left or right side of the road. This norm emerges through repeated and collective local interactions between agents, where everyone is expected to adopt the same behaviour in order to avoid conflicts and facilitate cooperation. 

Finally, personal norms, also known as moral or private norms, represent self-imposed internal norms. These norms ensure the agent's autonomy and reflect the agent's own values and principles, whether programmed or learned \cite{balke2013norms,tuomela1995importance,dignum1999autonomous,argente2020normative}. An example of this is the NPCD-Agent model \cite{peng2008extended}, an extension of the BDI model, which explicitly incorporates external and internal motivations for decision-making in agents. In that proposal, personal norms arise from the agent's internal and individual motivations. These motivations are not derived from the external environment or social obligations, but reflect internal preferences and goals that influence the agent's behaviour. Similarly, \cite{dechesne2012understanding} analyses how personal norms interact with legal and social norms in the context of smoking bans in public spaces. Classified agents make decisions based solely on their own internal principles, regardless of social expectations or legal regulations. The simulation results show that personal norms can play a crucial role in contexts where legal and social norms are not strong enough to ensure compliance.

\subsection{Life cycle of a norm in NMAS}

\begin{table}[t]
\centering
\caption{Classification of works usnig the live cycle phases: Creation, Diffusion and Adoption, and Internalization}
\label{table:normativity_phases}
\renewcommand{\arraystretch}{1.3} 
\begin{adjustbox}{max width=\textwidth}
\begin{tabular}{c p{5cm} p{5cm} p{5cm}}
\toprule
\textbf{Work} & \textbf{Phase 1: Creation} & \textbf{Phase 2: Diffusion and Adoption} & \textbf{Phase 3: Internalization} \\
\midrule
\cite{mukherjee2007emergence} & Norm emergence (explicitly referred to as "emergence") & Social learning (explicitly referred to as "social learning") & Agent interactions \\ 
\cite{savarimuthu2007mechanisms} & Norm emergence (mentioned in the context of the Ultimatum game) & Norm propagation (oblique and horizontal transmission, role model) & Internalization through group and personal normativity \\ 
\cite{mukherjee2008norm} & Norm emergence (explicitly mentioned) & Social learning and neighborhood influence (role model in neighbor networks) & Gradual adoption through learning in nearby neighbors \\ 
\cite{savarimuthu2007role} & Norm emergence (explicitly mentioned) & Norm propagation through role model in social networks and leadership & Behavioral alignment with role model norms through repeated interactions and normative advice \\ 
\cite{savarimuthu2009social} & Norm emergence (explicitly mentioned) & Norm propagation through peer-to-peer and distributed punishment supported by common knowledge & Internalization through repeated interactions and sanctions (when agents adopt norms to avoid punishments) \\ 
\cite{sen2009effects} & Norm emergence (implicitly through learning) & Norm propagation through proximity-based interaction and network topology & Adoption through continuous adjustment of behaviors after multiple network interactions \\ 
\cite{andrighetto2010complex} & Norm emergence (explicitly mentioned) & Norm propagation through normative recognition and shared normative beliefs & Adoption of normative beliefs incorporated into agents' decision-making \\ 
\cite{savarimuthu2011aspects} & Norm emergence (implicitly through active learning) & Propagation through observational learning, communication, and personal experience & Internalization through active learning (repeated interactions and sanctions or rewards) \\ 
\cite{brooks2011modeling} & Norm emergence (implicitly through coordination in repeated games) & Propagation through updating biases and reinforcing successful behavior & Adoption of behaviors through repeated bias adjustment based on successful interactions \\ 
\cite{ren2024emergence} & Creation \& representation & Spreading: communication (norm identification and influencing others) \& observation & Evaluation of norms (acceptance as personal norms) \\ 
\bottomrule
\end{tabular}
\end{adjustbox}
\end{table}

\begin{table}[t]
\centering
\caption{Classification of works based on normativity phases: Creation, Diffusion and Adoption, Internalization, and Forgetting}
\label{table:normativity_phases_forgetting}
\renewcommand{\arraystretch}{1.4} 
\begin{adjustbox}{max width=\textwidth}
\begin{tabular}{c p{3.7cm} p{3.7cm} p{3.7cm} p{3.7cm}}
\toprule
\textbf{Work} & \textbf{Phase 1: Creation} & \textbf{Phase 2: Diffusion and Adoption} & \textbf{Phase 3: Internalization} & \textbf{Phase 4: Forgetting} \\ 
\midrule
\cite{mahmoud2012efficient} & Norm emergence through compliance with dynamic sanctions & Propagation through repeated interaction and adaptive sanctioning & Behavior adjustment via experiential and adaptive punishment based on action history & Forgetting mechanism based on memory window \\ 
\cite{riveret2012probabilistic} & Norm emergence through probabilistic rule-based argumentation & Propagation via argumentation and probabilistic learning & Internalization by adjusting probabilities in defeasible logic & Forgetting mechanism through probability adjustment in reinforcement \\ 
\cite{yu2013emergence} & Norm emergence through collective learning and neighbor-based decisions & Diffusion via repeated interaction and aggregation of neighbor responses & Internalization through iterative behavior adjustment based on neighbor rewards & Decay of unreinforced preferences \\ 
\cite{yu2014collective} & Norm emergence through collective learning & Diffusion through continuous interaction and network reinforcement & Behavior adjustment through rewards from collective interaction & Decay of unreinforced preferences \\ 
\cite{shibusawa2014norm} & Norm emergence via weight influence in complex networks & Diffusion through propagation of influence weights in social networks & Internalization by adjusting based on cumulative influence weights in the network & Decay of influence based on interaction \\ 
\cite{beheshti2015cognitive} & Norm creation through cognitive (BDI) and social processes (GT interactions) & Diffusion via social learning and reinforcement & Internalization by adjusting beliefs, desires, and intentions (BDI) based on feedback from interactions & Decay in the value of unreinforced norms \\ 
\bottomrule
\end{tabular}
\end{adjustbox}
\end{table}

\begin{table}[htbp]
\centering
\caption{Classification of works based on normativity phases: Creation, Diffusion and Adoption, Internalization, and Transformation}
\label{table:normativity_phases_transformation}
\renewcommand{\arraystretch}{1.2} 
\begin{adjustbox}{max width=\textwidth}
\begin{tabular}{c p{3.7cm} p{3.7cm} p{3.7cm} p{3.7cm}}
\toprule
\textbf{Work} & \textbf{Phase 1: Creation} & \textbf{Phase 2: Diffusion and Adoption} & \textbf{Phase 3: Internalization} & \textbf{Phase 5: Transformation} \\
\midrule
\cite{savarimuthu2009norm} & Creation, emergence & Propagation & Identification & Reinforcement \\ 
\cite{campenni2009normal} & Norm emergence through normative recognition & Propagation via communication of normative beliefs and behavior observation & Internalization through the formation of normative beliefs guiding behavior & Innovation of norms due to barriers and external conditions \\ 
\cite{beheshti2014normative} & Norm creation via personal values and health norms & Diffusion through social networks and environment observation & Internalization through belief adjustment based on health values and social norms & Gradual adjustment in health-related beliefs \\ 
\cite{yu2015hierarchical} & Norm emergence through hierarchical learning with supervision from higher levels & Diffusion via hierarchical supervision and policy adjustments in networks & Internalization through Q-value adjustment based on supervisor information & Transformation through hierarchical feedback and elimination of subnorms \\ 
\cite{mahmoud2015establishing} & Norm emergence through interaction between agents with metanorms & Diffusion via enforcement of norms and metanorms in the agent network & Internalization through compliance with norms due to sanctions and metanorms & Adaptation of norms through behavior review \\ 
\cite{mashayekhi2016silk} & Norm emergence through recommendations from a central generator & Diffusion via suggested norms and experience sharing & Internalization through learning and adjustment to normative recommendations & Adjustment of norms based on accumulated experience \\ 
\cite{yu2016adaptive} & Norm emergence through supervised adaptive learning & Diffusion via social learning and supervision of neighboring agents’ behaviors & Internalization through continuous behavior policy updates based on rewards & Norm adjustment in response to feedback from adaptive learning \\ 
\cite{mahmoud2017establishing} & Norm emergence via use of metanorms and policy adaptation & Diffusion through interactions in specific topologies & Internalization through adaptation and learning of optimal behavior in the network & Behavior adjustment based on learning and interaction outcomes \\ 
\cite{macanovic2024signals} & Norm emergence through identity signaling & Diffusion via recognition of signals in conflict environments & Internalization through learning signals of group belonging & Norm change based on recognition and trust \\ 
\cite{ajmeri2018robust} & Norm emergence through contextual revelation and reasoning & Propagation through reasoning about shared context & Internalization via norm adjustment based on revealed social context & Norm refinement based on revealed context \\ 
\cite{abeywickrama2023emergence} & Actions (implicit norms) \& reinforcement learning & Topology \& neighbors observation & Not mentioned & Reinforcement learning \\ 
\cite{levy2023convention} & Norm emergence via action selection in congested environments & Diffusion through interaction with the environment and reward manipulation & Internalization by adjusting behaviors according to rewards manipulated by authority & Replacement of suboptimal norms through authority intervention \\ 
\cite{tzeng2021noe} & One social norm predefined & Identification & Not mentioned & Emotions serve as mental objects and an approach to sanctioning \\ 
\bottomrule
\end{tabular}
\end{adjustbox}
\end{table}

\begin{table}[htbp]
\centering
\caption{Summary of approaches to the life cycle of social norms in NMAS considering all the phases.}
\label{table:norm_lifecycle_mas}
\renewcommand{\arraystretch}{1.4} 
\begin{adjustbox}{max width=\textwidth}
\begin{tabular}{c p{3cm} p{3cm} p{3cm} p{3cm} p{3cm}}
\toprule
\textbf{Work} & \textbf{Phase 1: Creation} & \textbf{Phase 2: Diffusion and Adoption} & \textbf{Phase 3: Internalization} & \textbf{Phase 4: Forgetting} & \textbf{Phase 5: Transformation} \\ 
\midrule
\cite{hollander2011using} & Creation, emergence & Transmission & Recognition, application, acceptance, internalization & Forgetting & Evolution \\ 
\cite{mahmoud2016cooperation} & Norm emergence through resource control and peer punishment & Diffusion via punishments applied under resource constraints & Internalization through learning cooperative behavior via adaptive punishments & Gradual decay of unreinforced norms & Adjustment of punishments based on available resources \\ 
\cite{yang2016accelerating} & Norm emergence through heuristic hierarchical strategy (HHLS) & Diffusion via hierarchical supervision and instructions from supervisor agents & Internalization through continuous strategy updates based on supervisor rewards & Gradual decay of unreinforced norms & Strategy adjustments based on hierarchical feedback \\ 
\cite{hao2017efficient} & Norm emergence through heuristic learning and adjusted Q-learning & Propagation via collective learning in networks (FMQ and local/global exploration) & Internalization through updating Q-values and interaction frequencies in the network & Decay of unreinforced values & Norm adjustment based on accumulated performance and continuous learning \\ 
\cite{yu2017collective} & Norm emergence through collective learning and information diffusion & Diffusion via local interactions and information sharing among neighbors & Internalization through behavior adjustment based on cumulative neighbor rewards & Decay of unreinforced norms & Norm adjustment based on network conditions and received rewards \\ 
\cite{frantz2018modeling} & Creation & Transmission & Identification, internalization & Forgetting & Reinforcement \\ 
\cite{liu2021local} & Norm emergence through self-reinforcing substructures dissolved by incremental social instruments & Propagation via incremental links to unite substructures & Belief adjustment through reinforced interaction in the network & Decay in self-reinforcing substructures without recent interactions & Dissolution of substructures through incremental social instruments \\ 
\cite{morris2021norm} & Norm creation via synthesis and requests from participating agents & Diffusion via centralized propagation by synthesizing agents & Internalization through adoption of explicit norms in agents' internal representation & Forgetting stage is not considered & Evolution of norms through revisions proposed by agents \\ 
\cite{mashayekhi2022prosocial} & Norm generation & Norm sharing to incoming members & Not mentioned & Reinforcement Learning (decay) & Updating (norms may change when the environment changes) prosocial (guilt) \\ 
\cite{tzeng2022fleur} & Norm emergence through Social Value Orientation (SVO) & Diffusion via learning prosocial orientation in the network & Internalization by decision adjustment based on stable social preferences and values & Decay of less reinforced preferences & Norm adjustment based on rewards and experiences \\ 
\bottomrule
\end{tabular}
\end{adjustbox}
\end{table}

As discussed in Section~\ref{sec:live_cycle}, norms follow a life cycle that covers different phases. From an NMAS perspective, different works have presented models to simulate different stages of the life cycle of standards. A summary of the treatment of the life cycle of standards in NMAS can be found in the Tables~\ref{table:normativity_phases}, \ref{table:normativity_phases_forgetting}, \ref{table:normativity_phases_transformation}, and~\ref{table:norm_lifecycle_mas}. As can be seen, some works have focused on the first phases of the cycle (phases 1-2) where specific mechanisms for the creation, identification and emergence of standards are introduced. For example, in \cite{savarimuthu2011norm,savarimuthu2009norm} the authors present a normative agent model based on the hybrid approach. To this end, a life cycle for NMAS is proposed based on five phases: creation, identification, propagation, reinforcement and emergence. Norms can emerge through interaction and social learning or they can be centrally prescribed by agents who take the role of leaders. 

Other authors have presented NMAS models that cover a larger number of life cycle phases. Generally, these models tend to cover stages 1 to 4 of the life cycle. For example, the work presented in \cite{mahmoud2014review} develops a life cycle composed of five stages: creation, emergence, assimilation, internalisation and elimination. The authors use the concept of assimilation to control the evolution of norms in the system based on the evaluation of costs and benefits by actors. 

Finally, some proposals present a life cycle that covers a larger number of phases (phases 1-5). An example of this can be found in the work presented in \cite{hollander2011current,hollander2011using}. In that work, a life cycle for NMAS is presented based on ten stages: creation, transmission, recognition, application, acceptance, modification, internalisation, emergence, forgetting, and evolution. In that model, norms are created and transmitted until they are internalised by actors, leading to their emergence as established norms in society. Moreover, when a norm is no longer relevant, it is forgotten. Note that, the process of evolution covers the rest of the processes from creation to forgetting, which allows the system to adapt norms to the changing conditions of the social environment. Another interesting example can be found in \cite{frantz2018modeling}. This work proposes a global model for norm management in NMAS that integrates cohesively the aspects of creation, transmission, identification, internalisation, forgetting and reinforcement. To manage norm forgetting, a model of norm decay that alters the relevance of norms over time is presented. In that approach, norm emergence can follow two main perspectives: external design, in which human designers explicitly define and encode norms in agents; and autonomous innovation, where norms are spontaneously generated by agents from their interactions without external intervention.

\subsection{Factors influencing the emergence of norms in NMAS}
\label{sec:factors_emergence}

As have been show in Section~\ref{sec:emergency_factors}, there are a number of factors that are important in determining the efficiency and speed with which norms are established and consolidated in a society. These factors have also been addressed in NMAS for the implementation of mechanisms to enable the emergence of norms. In the following, the use of factors such as social topology, cognitive abilities of agents, propagation mechanisms, online and offline methods, emotions, and values will be reviewed within the scope of NMAS. Table~\ref{table:factors_NMAS} summarises the factors used by the proposals analysed in this systematic review. 

\subsubsection{Network Topology}

In the field of NMAS, different works have been proposed combining the different mechanisms related to network topology (see Section~\ref{sec:social_typology}). For example, in \cite{kittock1993emergent} the effect of network diameter on the emergence of norms in NMAS was studied. The results of the experiments showed that the diameter of the network was directly related to the convergence rate during the norm emergence process. Thus, a smaller diameter facilitates faster propagation of information and thus faster convergence. As the interaction radius in a circular network increases, the diameter decreases, which is associated with improvements in system performance. In \cite{mukherjee2008norm}, the effect of neighbourhood size on norm emergence was analysed. The authors showed that smaller neighbourhoods facilitated the emergence of norms, since, being in smaller groups, a greater number of interactions between agents took place. The effect of the clustering coefficient was shown in \cite{hao2017efficient}.

This coefficient measures the degree to which nodes tend to cluster. The results showed that, with a high clustering coefficient, such as small-world, the propagation of norms within local clusters was more efficient, taking advantage of short global paths for efficient diffusion. The work presented in \cite{sen2009effects} showed that the average path length also affected the emergence of norms. The authors showed that by using shorter paths, it was possible to increase the propagation speed of norms, facilitating their emergence. As in the previous example, small-world, scale-free networks were particularly effective for the emergence of norms. Another example of the analysis of network topology in NMAS can be found in \cite{savarimuthu2007mechanisms}. The authors highlight the importance of the degree distribution of the nodes in the network, emphasising that networks with heterogeneous degree distributions, such as scale-free networks, can significantly facilitate the emergence of norms. In these networks, the presence of hub nodes, which have a large number of connections, allows for a fast dissemination of information to other nodes, even in large networks. This is because these nodes act as central propagation points, reducing the need for multiple intermediaries in the communication process.

Some studies have analysed the influence of network structures on the emergence and propagation of norms from static and dynamic perspectives (\cite{sen2009effects,gelfand2024norm,du2024asymmetric,griffin2019consensus,savarimuthu2011norm}). On the one hand, in static networks, the network structures remain constant over time. In such networks, the choice of the overall structure has a significant impact on the evolution of the system and the convergence of norms. For example, in the work presented in \cite{pujol2006structure}, a scale-free topology is used to analyse the phenomenon of norm emergence. Such topologies are effective in norm propagation due to their power-law degree distribution and small diameter characteristics. Similarly, in \cite{mahmoud2017establishing}, a static network using a small-world topology is used for norm emergence. Small-world topologies are characterised by a small average distance between nodes, a high clustering coefficient and long-range links connecting distant nodes in the network. These characteristics make them more efficient than scale-free topologies as they allow for a better balance in the speed of norm propagation and the retention of local cohesion. 

On the other hand, dynamic networks allow for the creation, modification, or deletion of the relationships that occur between nodes. This feature makes them particularly useful for the realistic simulation of relationships and social interactions. Different studies have analysed the use of dynamic networks for the simulation of the emergence of norms in NMAS. For example, in \cite{griffiths2010norm}, the use of a tag-based cooperation model to analyse how norms emerge in dynamic networks is explored. In such a model, agents possess observable tags that act as social markers or signals. These tags represent social norms that are adopted by agents who share that tag, forming groups governed by those norms. Similarly, in \cite{savarimuthu2009norm} a model is proposed that uses the potential of dynamic networks to simulate a society of agents. The proposed process of norm emergence is based on learning and imitation of influential agents, who act as role models. By allowing agents to modify their social ties, the process of norm adoption in the network as a whole is accelerated. 

One problem that can arise in both static and dynamic network topologies is the emergence of local gated communities, or self-reinforcing substructures. This problem affects the propagation of global norms as it is constrained by the dynamics of the network itself. To address this problem, the Incremental Social Instruments (ISI) mechanism, designed to break down these substructures and facilitate the propagation of norms, is proposed in \cite{liu2021local}. The ISI mechanism acts by strategically adding links between agents from different self-reinforcing substructures, which breaks down these substructures and facilitates the propagation of norms globally. Through experiments on synthetic and real networks, ISI has been shown to be more efficient than static approaches, as it promotes the creation of social ties that connect previously isolated communities, thus accelerating the adoption of norms. 

\begin{table}[htbp]
\centering
\caption{Factors influencing the emergence of norms in NMAS.}
\label{table:factors_NMAS}
\renewcommand{\arraystretch}{1.5}
\begin{adjustbox}{width=\textwidth}
\begin{tabular}{p{5cm} p{5cm} p{5cm} p{5cm}}
\toprule
\textbf{Phase 1: Appearance} & \textbf{Phase 2: Propagation and adoption} & \textbf{Phase 3: Internalization} & \textbf{Forgetting  \textbackslash \newline Transformation} \\ 
\midrule
Normative assessor \cite{yang2016accelerating,frantz2014modelling} &  Social topology \cite{mukherjee2007emergence,savarimuthu2009norm,mukherjee2008norm,savarimuthu2007role,sen2009effects,hollander2011using,mahmoud2012efficient,yu2013emergence,yu2014collective,shibusawa2014norm,yang2016accelerating,yu2016adaptive,mahmoud2016cooperation,yu2017collective,hao2017efficient,mahmoud2017establishing,liu2021local,frantz2014modelling} &  Interaction learning \cite{savarimuthu2007mechanisms,savarimuthu2009norm,hollander2011using} &  Decision making \cite{savarimuthu2009social,savarimuthu2011aspects,brooks2011modeling,mahmoud2012efficient,shibusawa2014norm,mahmoud2015establishing,mashayekhi2016silk,yu2017collective,ajmeri2018robust,frantz2014modelling} \\
 Interaction learning \cite{mukherjee2007emergence,savarimuthu2007mechanisms,savarimuthu2009norm,mukherjee2008norm,savarimuthu2007role,savarimuthu2009social,andrighetto2010complex,sen2009effects,campenni2009normal,savarimuthu2011aspects,brooks2011modeling,hollander2011using,mahmoud2012efficient,yu2013emergence,beheshti2014normative,yu2014collective,shibusawa2014norm,mahmoud2015establishing,yu2015hierarchical,mashayekhi2016silk,yu2017collective,hao2017efficient,mahmoud2017establishing,morris2021norm,levy2023convention,liu2021local,tzeng2022fleur,ren2024emergence,macanovic2024signals,abeywickrama2023emergence,frantz2018modeling} &  Normative assessor \cite{savarimuthu2007mechanisms,levy2023convention} &  Learning \cite{savarimuthu2007mechanisms,andrighetto2010complex,campenni2009normal,frantz2014modelling} &  Other cognitive capacities \cite{mukherjee2007emergence,savarimuthu2007mechanisms,savarimuthu2009norm,mukherjee2008norm,savarimuthu2007role,campenni2009normal,yu2014collective,yu2015hierarchical,yang2016accelerating,yu2016adaptive,mahmoud2016cooperation,hao2017efficient,liu2021local,mashayekhi2022prosocial,ren2024emergence,macanovic2024signals,abeywickrama2023emergence} \\
 Learning \cite{riveret2012probabilistic,yu2013emergence,yu2014collective,beheshti2015cognitive,yu2015hierarchical,yu2016adaptive,yu2017collective,hao2017efficient,liu2021local} &  Model to follow \cite{savarimuthu2007mechanisms,savarimuthu2009norm,savarimuthu2007role,savarimuthu2009social,andrighetto2010complex} &  Other cognitive capacities \cite{riveret2012probabilistic,yu2013emergence,beheshti2015cognitive,morris2021norm,ren2024emergence,macanovic2024signals} & \\
 Other cognitive capacities \cite{mahmoud2016cooperation,ajmeri2018robust,mashayekhi2022prosocial} &  Interaction learning \cite{brooks2011modeling,hollander2011using,morris2021norm,ren2024emergence,macanovic2024signals} &  Values \cite{beheshti2014normative,tzeng2022fleur,mashayekhi2022prosocial,frantz2014modelling} & \\
 Values \cite{tzeng2022fleur} &  Punishment and reward \cite{mukherjee2007emergence,savarimuthu2009social,hollander2011using,mahmoud2012efficient,mahmoud2015establishing,mahmoud2016cooperation,mashayekhi2016silk,mahmoud2017establishing,ajmeri2018robust,levy2023convention,frantz2014modelling} & & \\
 Emotions \cite{tzeng2021noe} &  Observation \cite{andrighetto2010complex,campenni2009normal,savarimuthu2011aspects,macanovic2024signals} & & \\
&  Learning \cite{mukherjee2007emergence,mukherjee2008norm,sen2009effects,savarimuthu2011aspects,riveret2012probabilistic,yu2013emergence,beheshti2014normative,yu2014collective,shibusawa2014norm,beheshti2015cognitive,yu2015hierarchical,yang2016accelerating,yu2016adaptive,mashayekhi2016silk,yu2017collective,hao2017efficient,mahmoud2017establishing,ajmeri2018robust,liu2021local,tzeng2022fleur,tzeng2021noe,abeywickrama2023emergence,frantz2014modelling} & & \\
&  Emotions \& mood \cite{tzeng2021noe,mashayekhi2022prosocial} & & \\
&  Values \cite{tzeng2022fleur,mashayekhi2022prosocial} & & \\
\bottomrule
\end{tabular}
\end{adjustbox}
\end{table}

\subsubsection{Propagation Mechanisms}

\begin{table}[t]
\centering
\caption{Mechanisms for norm propagation in NMAS.}
\label{table:propagation}
\renewcommand{\arraystretch}{1.5} % Incrementa el espacio entre filas
\begin{adjustbox}{max width=\textwidth}
\begin{tabular}{p{3cm} p{3cm} p{3cm} p{3cm} p{3cm}}
\toprule
\textbf{Propagation Mechanism} & \textbf{Definition} & \textbf{Involved Actors} & \textbf{Propagation Method} & \textbf{Transmission Type} \\
\midrule
\textbf{Norm Advisor} & An entity acting as an authority providing expert guidance on prevailing norms. & Supervisory or control agents that collect feedback and adjust norms based on performance. & Direct and oblique diffusion through strategies and education. & Oblique \\ 
\textbf{Role Model} & Adoption of behaviors through imitation of reference figures. & Celebrities, community leaders, influencers. & Observation and imitation of behaviors. & Horizontal and Oblique \\ 
\textbf{Learning through Interaction} & Adoption of norms through daily social interaction. & Community members. & Observation of behaviors and outcomes of interactions. & Horizontal \\ 
\textbf{Incentive and Punishment} & Implementation of sanctions and rewards to ensure compliance with norms. & Sanctioning agents, incentive systems. & Punishment for non-compliance, rewards for compliance. & Horizontal and Oblique \\ 
\bottomrule
\end{tabular}
\end{adjustbox}
\end{table}

Propagation mechanisms (see Section~\ref{sec:propagation}) are essential for the cohesion between agents and the effective implementation of norms within NMAS. Table~\ref{table:propagation} shows a summary of the main propagation mechanisms used in the proposals analysed in this review: normative advisor, role model, learning through interactions, and incentives. There are different proposals that use the model of the normative advisor as a vehicle for the propagation of norms. For example, in \cite{savarimuthu2007mechanisms} they propose a norm advisor model as a mechanism to facilitate convergence towards a group norm. To do this, the policy advisor collects feedback from actors in the environment. Based on this information, the advisor adjusts the group norm to optimise collective outcomes and then communicates it to the agents. Similarly, in \cite{yang2016accelerating} it is shown how a supervisory agent acts as a normative advisor in a hierarchical model. This supervisor collects information from the interactions of subordinate agents and generates rules and suggestions to guide their strategies. The subordinate agents adjust their behaviour based on these guidelines, which accelerates the emergence of norms, as observed in coordination and anti-coordination scenarios, where this approach proves more efficient than non-hierarchical methods.

The use of role modelling has also been used in the NMAS field. For example, in \cite{morris2019norm} agents select their role models based on their success and reputation within society, and decide whether or not to accept their advice. This process of mutual influence reflects how the actions of highly influential agents tend to be replicated by the rest of the population. This method of behavioural propagation is known as horizontal transmission, although it can also be considered oblique transmission if the role models are multiple leaders in society. In \cite{savarimuthu2007mechanisms,savarimuthu2009norm}, the authors present a mechanism where the best performing agent, called the ``role model'', provides feedback to the other agents. They use an example based on the ultimatum game in which the role model agent helps agents facing difficulties during the course of the game. The agents, while maintaining their autonomy, can accept or ignore these suggestions, adapting their personal norms to improve their performance. On the other hand, \cite{franks2013manipulating} proposes the use of influencers as a vehicle for the propagation of norms. To do so, they used a ``lead-by-example'' perspective in which influencers played an active role in the dissemination and enforcement of these norms. These agents tried to change the behaviour of others by enforcing the norms, punishing offenders or rewarding those who comply. In \cite{sen2007emergence} they also used influencers to propagate norms. In that work the influencers were a set of agents with fixed strategies that guided the adoption of norms. To validate the approach, they used a scernario based on traffic norms. The results showed that influencers facilitated social learning and convergence towards shared norms. In \cite{haynes2017engineering} the concept of the influencing agent is also used. However, in this proposal the influencing agents had the ability to develop new norms and persuade other agents in their environment to follow them.  In \cite{shibusawa2014norm} an aggregative learning method based on the use of a degree of influence is proposed. This method facilitates the propagation of norms in complex networks by aggregating opinions and propagating the degree of influence. Agents adjust their behaviour not only according to the majority actions of their neighbours, but also considering the degree of consensus and influence of the most influential actors within the network, which contributes to the emergence of global rather than local norms. This approach allows the strategies of the most successful actors to propagate through the network, even in configurations with predominantly local interactions. Moreover, by dynamically integrating influence weights based on rewards and conformity, the method ensures that emerging norms reflect not only the local majority, but also strategies that maximise cohesion and coordination at the global level in the network.

Another well-studied mechanism in NMAS is interaction learning \cite{morris2019norm,chakrabarti2010emergence,hao2013achieving,hao2014multiagent,mukherjee2007emergence,sen2007emergence,villatoro2009topology,villatoro2011social,villatoro2013robust}. For example, in \cite{brooks2011modeling}, a model in which agents learn through one-to-one interactions in a coordination game is described. At the start of execution, action selection is based on a set of initial biases. This selection process is adjusted by considering the probabilities of success or failure in coordination. In this way, agents adapt their behaviour, facilitating convergence towards shared norms. In \cite{mukherjee2007emergence}, the use of interaction learning is also addressed. In that case, agents engage in stage games with opponents randomly selected from their neighbourhood or the general population. These interactions can follow either a uniform selection scheme, where all agents have equal probability of being selected, or a non-uniform scheme, which favours physical proximity between agents. The results showed the influence of local dynamics on the propagation of norms and the emergence of collective behaviour in heterogeneous societies. In contrast, other authors assume that the interactions that occur in the real world are not completely random but tend to be influenced by social proximity, such as between friends of friends or people in the same environment \cite{hassani2014dynamics,mungovan2011influence,peleteiro2014fostering,savarimuthu2007mechanisms,savarimuthu2011emergence,swarup2011model,yu2013emergence}. For instance, in \cite{hu2017achieving}, it is highlighted how learning through interactions drives the emergence of local conventions. Agents, without prior knowledge, engage in coordination games with neighbours and adjust their strategies based on rewards or penalties. Another type of interactions are weighted random interactions, which suggest that people are more likely to come into contact with specific individuals, thus better reflecting social dynamics. An example is the NMAS model proposed in \cite{hollander2011using} in which the process of adopting social norms is simulated. The model includes the transmission, enforcement and internalisation of norms, thus facilitating the formation of group consensus. Simulations showed that agents were able to detect and adopt norms through social interaction and information exchange in local networks, adjusting their beliefs and behaviours over time. Moreover, as the frequency and weighting of interactions increased, higher levels of convergence and faster adoption of norms in the group were observed.

Finally, incentives are one of the most commonly used mechanisms in NMAS to facilitate the propagation of norms. Incentives often include punishments to deter undesirable behaviours as well as rewards to encourage positive behaviours, are key to promoting compliance with norms \cite{castelfranchi1998principles,mahmoud2012efficient,mahmoud2015norm}. For instance, in \cite{savarimuthu2009social}, it uses a punishment-based system to encourage emergence, acceptance and compliance. To this end, it proposes the use of sanctioning agents that act on observed normative violations, facilitating behavioural change and social conformity. In \cite{lotzmann2013simulating} the authors propose the concept of norm invocations. Norm invocations represent the reactions of an actor when he or she feels offended or satisfied by the action of another. Negative invocations function as punishments, as agents issue them in the face of unacceptable behaviour, and those who receive them tend to reduce the likelihood of repeating that action in the future. Another interesting example can be found in \cite{mahmoud2014information}. The authors propose a model that simulates human motivation through the use of incentives. The model adjusts incentives over time according to the behaviour of the agents, thus maximising their effectiveness. In \cite{villatoro2011dynamic}, a dynamic punishment system is proposed, based on the number of offenders in society. This system allows sanctions to be adjusted according to fluctuations in normative behaviour. However, the use of agents to control violations is not always appropriate as there may be situations where agents may continue to violate norms because their violations are not sanctioned \cite{mahmoud2017establishing}. One way to avoid this phenomenon is presented in \cite{haynes2017engineering}. The authors use the concept of meta-standards. These meta-norms impose an obligation on agents to punish violators and discourage inaction in the face of transgressions, since there are also associated costs associated with not punishing. In a similar way, in \cite{mahmoud2015establishing}, meta-standards are used as an enabler for the emergence and enforcement of norms. However, limitations have also been identified in these models, such as the assumption that agents can observe all violations and punishments. To address this, some authors have resorted to the use of learning-based techniques. As an instance, in \cite{mahmoud2016cooperation} agents punish only transgressions that they directly observe, allowing for adaptive learning and limiting meta-punishment to more realistic situations. Similarly, in \cite{ajmeri2018robust}, agents adjust their behaviour and punish deviations from norms according to the context, making decisions based on the consequences of their actions, implementing meta-norms that punish both violators and non-punishers.

\subsubsection{Cognitive abilities}

One of the most common ways of simulating human social behaviour in NMAS is through the use of metaphors based on cognitive abilities such as observation, learning or reasoning skills. Observation is an active skill that allows agents to adjust their behaviour by observing the decisions of others in their immediate environment. This process facilitates convergence towards common norms without the need for explicit communication between agents \cite{epstein2001learning,savarimuthu2011norm}. Through observation, agents learn from each other's actions and adapt their behaviours, which accelerates the formation and adoption of norms within the group. Agents can develop their observation skills in different ways, either through learning by imitation, collective learning or direct observation. One of the ways in which observation materialises is through imitation, where agents replicate successful or high status behaviours within the group \cite{savarimuthu2011norm}. For example, in \cite{andrighetto2010complex,campenni2009normal} a norm recognition module called EMIL-A is proposed. In that model, agents identify the existence of a norm by interpreting observed behaviours or through the use of communication. This identification process generates a normative belief in the agent's knowledge base. These normative beliefs allow agents to know and follow socially accepted behaviours and to transmit this normative knowledge to other agents. 

An important aspect in the simulation of the observation process is the agents' ability to access information. In general, two types of proposals can be distinguished: those using a global observation approach, where agents observe all the actions of others; and those using a local observation approach, where each agent only has information about a subset of agents. For example, in \cite{hao2017dynamics}, the influence of the scope of observation on the emergence and propagation of norms is analysed. Simulation results showed that agents using a global observation model were able to converge to optimal policies faster than those using a local observation level. In \cite{yu2013emergence,yu2017collective} the authors use a local observation model for norm emergence using social learning. In this approach, agents had no direct access to information about the actions or outcomes of their neighbours. However, through the aggregation of decisions based on local experience, agents were able to learn and transmit norms of social behaviour. Similarly, the use of local observation to simulate the emergence of norms through the use of learning was analysed in \cite{mukherjee2008norm,morris2019norm}. Agents acquired private knowledge derived from their interactions within their neighbourhood. The results showed that, despite not having global knowledge, agents were able to reach global consensuses about behaviour by observing their neighbours.

Another cognitive ability commonly used for the simulation of norm emergence in NMAS is learning. One of the classic approaches is imitation learning, where agents replicate the behaviours observed in other agents in their environment. For instance, in \cite{epstein2001learning} agents learn by means of the ``Best Reply to Adaptive Sample Evidence'' strategy. This strategy is based on two key aspects: determining which norm to adopt by observing others and deciding how much cognitive energy to invest in it. The model is able to explain how consensus on normative behaviour is produced at the local level, highlighting how the evolution of norms reduces individual effort as they consolidate. It also offers an explanation for the diversity of behavioural norms in global settings. However, imitation learning has certain limitations in simulating the coexistence of multiple norms within a single system, which may restrict the diversity and flexibility of normative behaviours in complex environments \cite{campenni2009normal}. An alternative to mitigate this effect is the use of reinforcement learning \cite{chakrabarti2010emergence,hao2015heuristic,villatoro2009topology,yu2014collective,hao2013achieving,hao2014multiagent,mashayekhi2016silk}. As an example, in \cite{savarimuthu2010norm,savarimuthu2010obligation}, agents use sanctions and rewards to infer a set of norms that define their behaviour within society. Norm learning takes place autonomously, allowing agents to adapt dynamically to normative changes in their environment.  Similarly, in \cite{hao2015heuristic}, a networked collective learning framework is presented to facilitate the emergence of social norms through local interactions. The authors use a model based on reinforcement learning, where agents update their decisions based on past rewards. To do so, they use the Frequency Maximum Q-value heuristic, which allows prioritising actions with better historical outcomes. Another example related to collective reinforcement learning is presented in \cite{yu2014collective}. In that approach, each agent interacts with its neighbours and updates its decisions based on the rewards obtained during these interactions. However, this method incorporates collective aggregation methods, such as ensemble learning, to combine the optimal actions towards each neighbour into a final action. This process allows to simulate collective decision-making in humans, where multiple alternatives are usually evaluated before reaching a consensus. 

A common element in reinforcement learning is the use of memory to store past events. For instance, in \cite{villatoro2009topology} the authors use past events to identify norms of social behaviour. The results showed that agents' memory capacity influenced the speed at which they make decisions and the speed of convergence to a norm. Different alternatives have been proposed to mitigate the intensive use of memory. In \cite{riveret2012probabilistic} a reinforcement learning model is proposed in which a probability is associated with the norm that reflects its degree of internalisation. If the probability of a norm is 1, it is considered fully internalised, while a probability of 0 indicates that the norm is deactivated in the agent's mental profile. This allows the use of memory about past actions to be minimised. Other authors propose that the learning capacity depends exclusively on the current interaction, i.e. without storing any history of past interactions \cite{mukherjee2007emergence,mukherjee2008norm,villatoro2013robust,sen2007emergence}. As an example, a prosocially oriented reinforcement learning model is defined in  \cite{mashayekhi2022prosocial} in which past interactions are not considered. Agents seek to maximise the welfare of others by proactively avoiding unfair or unequal situations. To this end, agents dynamically adopt a set of social norms to minimise inequity. This system was evaluated in traffic simulations and was found to be efficient, equitable and welfare-promoting. Similarly, a model that learns through immediate interactions is used in \cite{levy2023convention}. In this way, agents structure their behaviour with the aim of maximising rewards in environments with high demand or competition for resources. 

There are different approaches to deal with reinforcement learning. In \cite{yu2013emergence} a Q-learning algorithm is proposed for the emergence of norms through interactions with neighbours. Agents used sanctions and rewards to adjust their strategies for following traffic norms. Another example of this learning mechanism is proposed in \cite{sen2007emergence}, in which it is used as a central algorithm in a social learning framework to study the evolution of norms among agents interacting on a recurrent basis. In this scenario, agents engage in coordination games and social dilemmas, adjusting their decisions based on the rewards obtained in individual interactions. This approach allows agents to converge towards emergent social norms, such as resolving the ‘give way’ dilemma at intersections. Furthermore, learning is fully distributed, based exclusively on private experiences. Similarly, a reinforcement learning algorithm based on Win or Learn Fast - Policy Hill Climbing (WoLF-PHC) is used in \cite{yu2016adaptive}. WoLF-PHC introduces an adaptive mechanism into an agent's learning rate, adjusting its behaviour depending on whether it is ``winning'' or ``losing'' \cite{mukherjee2008norm,yu2016adaptive}. Agents dynamically adapt their learning and exploration rates based on their performance against optimal strategies. This approach allows agents to adjust their behaviour more quickly when their strategies are not effective, and to be more conservative when they are performing well \cite{hu2017achieving}. Other work uses game theory for rule emergence. In \cite{mukherjee2008norm}, the authors use Fictitious Play, a method in game theory where individuals assume that their opponents play fixed strategies based on the frequency of their past actions. They used a scenario in which agents had to decide on a common side of the road at busy intersections. Using this methodology, the agents were able to gradually adapt and agree on a set of norms. Furthermore, agents were able to form expectations about each other's behaviour and adjust their own actions accordingly. However, WoLF-PHC and Fictitious Play are less efficient in distributed agent environments as demonstrated in the experiment conducted in \cite{mukherjee2008norm}. The authors analysed the ability of these two algorithms versus Q-learning to reach consensus in rule emergence processes. The results showed that by using Q-learning, agents reached consensus more quickly. This is attributed to its ability to balance the exploration of new strategies and the optimisation of decisions based on accumulated experience. In addition, smaller neighbourhoods were found to accelerate local coordination through Q-learning, favouring the emergence of norms more efficiently.

Learning in norm emergence simulation environments can also be supervised. Under this paradigm, agents learn a catalogue of norms that has to be previously formalised and structured. This type of learning facilitates a rapid adoption of predefined norms and promotes cohesion and uniformity within a group \cite{sen2007emergence,bicchieri2005grammar}. This has been demonstrated in studies of norm emergence through coordination games, where agents are guided towards specific normative behaviours \cite{shoham1992emergent}. For example, in \cite{sen2007emergence}, a scenario is used where agents learn traffic norms. They use a supervised learning model in which correct actions are rewarded, while incorrect behaviour is penalised. In contrast, unsupervised learning allows agents to discover patterns or structures in their experiences without explicit guidance, being particularly valuable in hierarchical \cite{yang2016accelerating} contexts. In these scenarios, individuals can identify norms that operate at different levels of social organisation. As an example, in \cite{yu2015hierarchical}, unsupervised learning is explored in a scenario in which agents decide in a coordinated way on acceptable social behaviour. A hierarchical structure inspired by human social organisations is used. This implies that the system represents different legislative layers that are responsible for balancing the interactions between agents.  Another example is presented in \cite{yang2016accelerating}, where the hierarchically heuristic learning strategy has been taken into account, where subordinate agents report their experiences to supervisors, who generate norms or suggestions to guide their strategies. This combines local learning and centralised supervision. This hierarchical framework allows agents to adjust their behaviours based on both their own experience and instructions from supervisors, which is crucial when norms are not clearly defined and must be inferred from observation and experience.

Another aspect to take into account in the learning process is the individuals involved in the learning process: peer or collective learning. In peer learning an agent adjusts its behaviour based on a single interaction with another agent during a specific round \cite{yu2013emergence,sen2007emergence,savarimuthu2011emergence,villatoro2013robust}. In contrast, collective learning allows an agent to adjust its behaviour by considering interactions with several agents in the same round \cite{hao2017efficient}. Collective learning allows a faster convergence towards shared norms than peer learning \cite{yu2013emergence}. This is because collective learning provides a more complete view of the social environment, allowing actors to make more informed and effective decisions. As an alternative example, a heuristic model for collective learning is proposed in \cite{hao2015heuristic}. The authors present a small-world scenario in which agents interact locally with several neighbours at a time. Agents adjust their strategies based on a record of past performance and frequencies of success. This allows them to coordinate and facilitate the emergence of social norms efficiently.

Finally, agent models generally use different approaches to simulate the agent's reasoning process. In an agent, the reasoning process allows it to decide on the strategies or actions it can take based on its knowledge base. In the emergence of norms in NMAS, there are different proposals for dealing with the agent's reasoning process that can be grouped into three basic levels of reasoning: low, medium and high. Low-level reasoning processes tend to use simple strategies to select actions, such as fixed choice or random selection among several predefined options \cite{hao2013achieving,mukherjee2008norm,sen2007emergence}. An agent model with a low level of reasoning is proposed in \cite{hao2013achieving}, the emergence of norms in stylised games such as the prisoner's dilemma. Agents use action selection strategies based on reinforcement learning, employing a mechanism that balances the exploitation of known actions and the exploration of new options. Strategies are adaptively updated between individual and social schemas according to past performance, allowing to coordinate actions, avoid exploitations and achieve socially optimal norms in a stable and efficient way in the system. In \cite{mukherjee2008norm}, agents, through a context of distributed interactions on a spatial grid only interact with near neighbours, using action selection strategies based on social learning, allowing agents to develop behavioural policies by observing the actions of their opponents in stage games. 

Agents with average cognitive ability have skills that allow them to choose among multiple options depending on the context or to employ learning algorithms to determine the best action based on previous experiences \cite{villatoro2013robust,savarimuthu2009social,brooks2011modeling}. For example, in \cite{ghorbani2017self}, a scenario is presented in which agents evaluate their energy level to decide whether to continue with their current strategy, innovate or imitate more successful neighbours. This approach allows them to adjust their behaviour to the environment, favouring the emergence of collective norms and the sustainability of shared resources. In \cite{villatoro2013robust} strategies such as observation and rewiring allow agents to influence the emergence and consolidation of norms are analysed. Observation reinforces prevailing normative behaviours by gathering information from neighbouring nodes, while rewiring enables the strategic modification of connections to interact with like-minded agents. These strategies are especially effective in complex networks, such as scale-free networks, by resolving local arrangements resistant to change and facilitating convergence towards global norms.

Agents with a high cognitive capacity are able to engage in deep reasoning about their environment, the prevailing norms and possible actions to take. This advanced level of decision-making enables them not only to follow norms, but also to actively assess their relevance to the circumstances at hand \cite{conte1995cognitive,verhagen2000norm}. In contrast to agents with average cognitive ability, they are distinguished by their ability to reason autonomously, to evaluate norms critically and to adapt their normative behaviour according to the context. For instance, in \cite{ajmeri2018robust}, agents use their advanced cognitive abilities to decide whether to comply with or deviate from social norms according to the context. These agents evaluate factors such as social relatedness and the urgency of actions, revealing contextual information when necessary to justify deviations and foster social cohesion. This level of reasoning allows them to adapt norms to dynamic situations and maximise social experience, standing out for their autonomy and flexibility in decision-making. On the other hand, in \cite{yu2017collective}, agents exhibit advanced cognitive capabilities through a collective learning framework, where they make decisions based on repeated interactions with their neighbours and using reinforcement learning strategies, such as Q-learning. These final decisions are made using opinion aggregation methods that integrate information from the environment, such as the structure of the network and the historical performance of neighbours. Additionally, agents engage in social learning processes, exchanging information to evaluate and adjust norms according to the context, which highlights their ability to make informed and adaptive decisions in complex networks. 

\subsubsection{Online-offline methods}
In the literature it is possible to identify two main approaches to norm synthesis in agent societies: online and offline methods. Online norm synthesis refers to the process in which norms emerge and are adjusted during the execution of the system. In this way, norms adapt to new situations as agents interact \cite{morris2019norm}. For example, in \cite{franks2013learning}, agents are introduced at strategic positions within the network, i.e. positions selected based on topological metrics such as centrality and connectivity. During the execution of the system the agents have a power to significantly influence the normative evolution of society. To this end, agents adopt fixed strategies that facilitate the propagation of norms or conventions. This approach was evaluated in a model of linguistic coordination, where Influencer Agents were able to guide the population towards a dominant convention more efficiently.

In contrast, approaches based on offline methods seek to anticipate possible scenarios by defining norms before the system is operational. This approach is more appropriate in contexts where most system states can be anticipated and interactions between agents are relatively predictable. For example, in \cite{morales2018off} different offline methods are defined that use simulations of interdependent conflict games to generate rules that are then enforced in the system. However, this approach can be limited in dynamic environments, as predefined rules may not adapt to unforeseen circumstances, which could lead to the need for adjustments or even a complete reconfiguration of the system.

One of the most important challenges in both approaches, but particularly relevant in the online method, is the management and resolution of conflicts between different norms or behavioural expectations. In complex systems, where actors may play multiple roles and face conflicting situations, it is essential to have mechanisms that allow for dynamic adjustment of norms in order to maintain the cohesion of the system.  In \cite{corapi2011normative,athakravi2013handling}, frameworks based on inductive logic programming have been developed that facilitate the continuous adaptation of norms and the resolution of normative conflicts, ensuring greater flexibility in changing environments.

\subsubsection{ Emotions in NMAS}
Emotions have been increasingly recognised as a key mechanism in the emergence of norms, acting as a link between individual experience and social regulation \cite{van2017emotions}. Emotions play a role in how the actions of others are perceived and evaluated \cite{van2010interpersonal}. For example, anger may arise in response to the observation of the violation of a social norm. This negative response serves as an emotional punishment and encourages compliance with norms \cite{haidt2003moral,fehr2002altruistic}. An example of the use of emotions in agents can be found in \cite{von2014emotion}. In that proposal, each agent has an emotional state function that reflects its level of satisfaction or dissatisfaction, depending on the interactions and the results of the actions. When agents experience negative emotions, such as anger at an action perceived as unfair, the emotion-based norm emergence mechanism is activated. This process generates norms that discourage the actions that triggered the negative emotions, promoting fairer and more cooperative behaviour. Similarly, a normative emotional agent model is presented in \cite{conte2014minding}. The agent is able to react by using negative emotions to actions that violate social norms, such as telling a lie. In \cite{von2014emotion} each agent possesses an emotional state function that reflects its level of satisfaction or dissatisfaction, depending on the interactions and outcomes of the actions. When agents experience negative emotions, such as anger at an action perceived as unfair, the emotion-based norm emergence mechanism is activated. This process generates norms that discourage the actions that triggered the negative emotions, promoting fairer and more cooperative behaviour. In this way, individual emotions influence normative regulation at the macro level within the agent society. 

In addition to reporting normative violations, emotions also play a significant role in the internalisation and propagation of norms within a social group. Different works have studied the use of emotions in the processes of internalisation and propagation of norms \cite{von2006my,staller2001introducing}. In \cite{fix2006emotion}, the use of emotions to model the emergence and maintenance of social norms in NMAS is proposed. Emotions are used to reinforce and preserve norms, establish dynamic and flexible control, and improve the representation of social interactions. In \cite{tzeng2021noe}, the authors present an agent architecture (Noe) that integrates decision-making with normative reasoning and emotions. Noe agents evaluate the environment by considering the emotions expressed by others, their mental states and the possible consequences of their actions. Experiments in simulated environments show that agents who incorporate emotions cooperate better and comply with norms more effectively than those who do not. Experimental results show that agent societies in Noe achieve greater normative cohesion, improve social welfare and provide a more positive experience compared to societies based only on sanctions or anarchy, which reinforces the robustness of norms in the system. While not all NMAS require emotional agents, the inclusion of emotions is especially relevant in contexts where agents interact with human users \cite{gratch2005evaluating}. The incorporation of emotions in the process of norm emergence seeks to contribute to the creation of more equitable and harmonious systems by addressing actions that provoke negative emotions in other \cite{beale2009affective} agents. This approach allows norms to be not only externally imposed, but also internalised by individuals, which promotes greater social cohesion and behaviours more aligned with group expectations.

\subsubsection{Values in NMAS}
From a philosophical perspective, values are understood as fundamental principles that guide human decisions, reflecting ethical and cultural ideals about what is considered desirable or right. In the field of social interaction, values act as a mechanism for shaping and balancing social dynamics \cite{dignum2019responsible}. In NMAS values have also been applied as a factor for the emergence of norms \cite{mosca2021elvira,szekely2021evidence,winikoff2021bad}. In \cite{tzeng2022fleur} a value-based model is proposed for a pandemic scenario in which agents with different social value orientations coexist based on the dimensions: altruism, prosociality, individualism, competitiveness. Agents must decide whether or not to comply with a norm of wearing a face mask in public settings. In making the decision, agents consider their personal preferences and social considerations. In \cite{tzeng2022noe}, agents' underlying values are reflected in the evaluation of social norms during interactions in a simulated environment. Although not explicitly modelled, it is inferred that normative compliance decisions are influenced by individual and social preferences, suggesting the importance of values in the adoption and stability of emerging norms. Moreover, in \cite{mercuur2019value}, values are modelled as guiding principles for agents' behaviour, specifically in the context of the ultimatum game. It is emphasised that values, such as equity and wealth, directly influence agents' decisions by determining the relative weight of these priorities. Values are treated as a static component in decision-making, which consistently guides agents' behaviour. In \cite{ajmeri2020elessar}, the authors propose a model in which agents consider both their own preferences and the values of others. In \cite{bench2017norms,serramia2018exploiting}, it is suggested that agents use value-based arguments to decide their actions, whether to comply with or violate norms. In \cite{cranefield2017no}, it addresses how values can influence plan selection in agents, prioritising those plans that are most consistent with the agent's values to achieve a goal, although the interaction between values, goals and norms is not explored in depth. In \cite{szabo2020understanding} an agent model is proposed that combines values and norms to simulate behaviour aligned with human expectations. Similarly, in \cite{serramia2018moral}, we analyse how moral values can guide normative decision-making, prioritising norms that promote society's preferred values.  In \cite{kayal2018automatic}, a model for resolving normative conflicts in assistive technologies is presented, using value profiles to prioritise norms that promote key values such as privacy and security. The model evaluates both normative trade-offs and users' general preferences, selecting the norm that best aligns with their values.

Another interesting proposal on the application of values in NMAS can be found in \cite{mashayekhi2022prosocial}. The authors develop a model in which values are manifested in a road intersection scenario where agents make prosocial decisions based on inequity aversion. Agents consider both their own and others' costs, adjusting their behaviour to reduce disparities, such as yielding to vehicles that have waited longer. Similarly, in \cite{tzeng2021noe}, values are manifested in a grocery shop queuing scenario. Agents have to decide whether to respect the norm of waiting in line or to violate it by skipping the line to get food faster. Agents' decisions are influenced by a combination of social values and emotional states. For example, an agent who values fairness may decide to wait his or her turn to avoid negative emotions, such as guilt or shame, that might arise from violating the norm. 

\subsection{Discussion}
The models and approaches discussed in this paper demonstrate the complexity and diversity inherent in the process of norm emergence in multi-agent systems. From our analysis, several key issues have been identified that require attention in order to advance the design and understanding of these systems.

The models and approaches discussed in this paper highlight the complexity and diversity inherent in the process of rule emergence in multi-agent systems. From our analysis, several key aspects have been identified that require attention in order to advance the design and understanding of these systems, the first of which is the distinction between prescriptive and emergent approaches. While prescriptive models, characterised by the centralised imposition of norms, offer a formal and stable structure, their rigidity makes them less suitable in dynamic environments \cite{savarimuthu2011emergence,criado2010normative,mukherjee2007emergence}. On the other hand, emergent approaches are notable for their adaptability, as norms emerge spontaneously from interactions between agents \cite{savarimuthu2011aspects,yu2013emergence,yu2014collective}. However, this flexibility can also lead to regulatory fragmentation, making overall coordination within the system difficult. Combining both paradigms in hybrid models, as suggested in recent work, allows the advantages of both approaches to be exploited, integrating centralised control with contextual adaptive capacity \cite{macanovic2024signals,morris2021norm,yu2015hierarchical}.

Another critical element is the life cycle of norms, from their creation and propagation to their internalisation and eventual forgetting. While some models focus exclusively on the initial stages \cite{savarimuthu2009norm,sen2009effects,mukherjee2008norm}, others offer a more holistic view by including later phases \cite{hollander2011using, frantz2018modeling,yu2017collective}. Internalisation, in particular, emerges as a crucial phase that has received little attention in the current literature. To ensure normative stability in complex systems, it is essential to develop models that include both the effective assimilation of norms and the elimination of those that are no longer functional.

Structural factors also play a key role in the emergence and spread of norms. Network topology significantly influences regulatory diffusion and stability. For example, loosely scaled networks, with their hubs and short paths, are effective for rapid propagation, but are vulnerable to fragmentation if key nodes are removed \cite{sen2009effects, hao2017efficient}. In contrast, more structured networks, such as lattice networks, offer local cohesion but limit global diffusion. Dynamic networks present themselves as a promising solution, as they allow for adjustments in the interactions between agents, facilitating both the propagation and the coexistence and evolution of local norms. Designing robust networks that balance connectivity, resilience and adaptability is a key direction for future research.

In terms of propagation mechanisms, the diversity of approaches makes it possible to adjust actors' behaviour according to their context. Strategies such as the normative advisor and role model are notable for their ability to influence through both formal authority and informal persuasion. Interaction-based learning also fosters more adaptive propagation \cite{brooks2011modeling,mukherjee2007emergence}, although it faces challenges such as the emergence of local sub-norms that inhibit global adoption \cite{hao2014multiagent, hollander2011using}. Finally, rewards and punishments represent a formal mechanism to ensure regulatory compliance, although they must be carefully designed to avoid adverse effects such as rejection or loss of intrinsic motivation \cite{savarimuthu2009social,ajmeri2018robust,mahmoud2015establishing}.

In addition to structural aspects, cognitive and emotional factors play a crucial role in the emergence of norms. From a cognitive perspective, agents' ability to observe, imitate and actively learn allows them to adjust their behaviour to the dynamics of the system. While agents with limited skills tend to conform through simple processes of imitation \cite{campenni2009normal,andrighetto2010complex}, those with greater cognitive development evaluate the relevance of norms in different contexts, promoting flexibility and dynamism \cite{yu2013emergence,yang2016accelerating,levy2023convention}. This balance between simplicity and complexity poses a challenge for designing regulatory systems that are both efficient and adaptive.

In the emotional domain, emotions act as catalysts in the regulation of normative behaviour. Responses such as guilt, shame or anger not only reinforce conformity, but also promote social cohesion and normative stability \cite{tzeng2021noe,von2014emotion}. Incorporating emotional mechanisms into NMAS can foster more effective internalisation of norms, resulting in more harmonious and equitable systems.

Shared values within a community act as a normative framework that reinforces the emergence and adoption of norms. When values and norms are aligned, conformity is facilitated and group cohesion is strengthened. However, conflicts between values can destabilise norms, generating tensions within the social network. This finding underlines the importance of adopting a holistic perspective that considers the interaction between structural, cognitive, emotional and cultural factors in the design of normative \cite{tzeng2021noe,mashayekhi2022prosocial} models.

Despite the progress identified, several open challenges remain that limit the development and application of normative models in multi-agent systems. First, the integration of multiple regulatory approaches into a single operational model remains a significant challenge, especially in dynamic contexts where norms must constantly adapt to new conditions. In addition, assessing the effectiveness of standards and mechanisms to measure their impact over time requires more robust methodological tools.

Another important challenge is the balance between computational efficiency and model complexity. Designing agents that integrate structural, cognitive, emotional and cultural factors without compromising system performance is a critical task. Also, the interaction between norms and other cognitive processes, such as decision-making and learning, needs to be explored in more detail to understand how these interdependencies affect collective behaviour.

Finally, designing multi-agent systems that can operate in multicultural environments poses additional challenges. The coexistence of divergent values and norms in the same social network requires innovative approaches to manage normative conflicts and promote cohesion in heterogeneous contexts. These challenges highlight the need for interdisciplinary collaboration to advance the field.

\section{Conclusions}
\label{section:conclusions}
This paper has presented a comprehensive review of the state of the art in norm emergence in NMAS. We have analysed the main factors involved in norm emergence including different normative approaches, propagation mechanisms, and structural, cognitive, emotional and value factors. By exploring different approaches at the structural and cognitive level, we have identified critical elements that influence how norms emerge, propagate and stabilise in different environments. This holistic perspective highlights the multifaceted nature of normative systems and the need to develop integrative models that address both theoretical and practical considerations.  

The results of this review have shown the main lines of current research in the field of standards emergence in NMAS. The two main approaches to standards emergence have been analysed: prescriptive approach and emergent approach. The importance of the proper development of the life cycle of standards, from their creation and dissemination to their internalisation and eventual obsolescence, has also been analysed. It has been shown how the structure of social networks, represented by factors such as clustering coefficient, centrality and weak ties, plays a crucial role in the propagation and consolidation of norms. Free-scale networks, known for their capacity for rapid diffusion, are vulnerable to fragmentation in the face of targeted disruptions. Dynamic networks, capable of evolving patterns of interaction between agents, represent a promising area for future research, as they facilitate both normative adaptability and resilience.

The main factors that facilitate the intenalisation of norms such as observation, social learning and imitation have been developed. On the other hand, the propagation mechanisms examined reveal a variety of strategies to influence agents' behaviour. From influences based on authority and peer persuasion to reinforcement through rewards and punishments, each mechanism has distinctive strengths and limitations. In addition to structural and procedural factors, cognitive and emotional dimensions have a significant impact on normative emergence. Emotions such as guilt, shame and anger act as powerful tools to reinforce norms and promote social cohesion. 

Despite this progress, a number of challenges remain. The integration of diverse regulatory approaches within a cohesive framework remains a major obstacle, especially in environments that require constant adaptation. Methodologies for assessing the long-term effectiveness of standards, including their life-cycle dynamics, need further development to ensure robust evaluation capabilities. Computational efficiency poses another challenge, as models incorporating complex structural, cognitive and emotional factors must remain scalable and practical for real-world applications.

\bibliographystyle{plain}
\bibliography{bibliografia}

\end{document}